%% file: n4-correlator.tex
\documentclass[a4paper,11pt]{article}
\pdfoutput=1 

\usepackage{jheppub} 

\usepackage[T1]{fontenc} 

\usepackage{graphicx} 
\usepackage{caption} 
\usepackage{amsmath} 
\usepackage{verbatim} 
\usepackage{mathtools}
\usepackage{url}
\usepackage{setspace}
\usepackage{breqn}
\usepackage{subcaption}
\usepackage{float}
\usepackage{leftidx}
\usepackage{bm}
\usepackage[utf8]{inputenc}
\allowdisplaybreaks

\usepackage{amsmath,amssymb,braket,cancel,nicefrac,physics,tensor,slashed}
\usepackage{color,empheq,enumitem,forloop,graphicx,microtype,subcaption,verbatim,wrapfig,pdfpages,titlesec}
\usepackage{array,booktabs,multicol,multirow,tabularx} 

\def\Circlearrowleft{\ensuremath{%
  \rotatebox[origin=c]{180}{$\circlearrowleft$}}}
\def\Circlearrowright{\ensuremath{%
  \rotatebox[origin=c]{180}{$\circlearrowright$}}}

\def\be{\begin{equation}}
\def\ee{\end{equation}}
\def\ba{\begin{eqnarray}}
\def\ea{\end{eqnarray}}
\def\nl{\nonumber\\}
\def\zb{\bar{z}}
\def\a{\alpha}
\def\ab{\bar{\alpha}}
\def\hb{\bar{h}}
\def\l{\langle}
\def\r{\rangle}
\def\OO{\mathcal{O}}
\def\GG{\mathcal{G}}
\def\HH{H}

\def\dDisc{\text{dDisc}}
\def\D{\mathcal{D}}
\newcommand{\lsim}{\mathrel{\hbox{\rlap{\lower.55ex \hbox{$\sim$}} \kern-.3em \raise.4ex \hbox{$<$}}}}
\newcommand{\gsim}{\mathrel{\hbox{\rlap{\lower.55ex \hbox{$\sim$}} \kern-.3em \raise.4ex \hbox{$>$}}}}

\title{All Tree-Level Correlators in AdS${}_5\times$S${}_5$ Supergravity:\\ Hidden Ten-Dimensional Conformal Symmetry}

\author[a]{Simon Caron-Huot,}
\author[a]{Anh-Khoi Trinh}

\affiliation[a]{Department of Physics, McGill University, 3600 Rue University, Montr\'eal, QC Canada H3A 2T8}

\emailAdd{schuot@physics.mcgill.ca}
\emailAdd{anh-khoi.trinh@mail.mcgill.ca}

\abstract{
We study correlators of four protected (half-BPS) operators in strongly coupled supersymmetric Yang-Mills theory.
These are dual to tree-level supergravity amplitudes on AdS${}_5\times$S${}_5$ for various spherical harmonics on the five-sphere.
We use conformal field theory methods,
in particular a recently obtained Lorentzian inversion formula, to analytically bootstrap these correlators.
The extracted $1/N^2$ double-trace anomalous dimensions confirm a simple pattern
recently conjectured by Aprile, Drummond, Heslop and Paul.
We explain this pattern by an unexpected ten-dimensional conformal symmetry which appears to be enjoyed by
tree-level supergravity (or a suitable subsector of it).
The symmetry combines all spherical harmonics into a single ten-dimensional object,
and yields compact expressions for the leading logarithmic part of any half-BPS correlator at each loop order.
}

\begin{document} 
\maketitle
\flushbottom

\section{Introduction}

Among strongly coupled quantum systems, a distinguished class are dual
to weakly coupled gravitational theories and are therefore efficiently tackled
by the gauge-gravity correspondence.
Models in this class are characterized by a large-$N$ expansion and a sparse spectrum of local operators:
all single-trace primary operators with small scaling dimension have spin two or less \cite{Heemskerk:2009pn}.
While the gravity side of the correspondence typically remains more straightforward to use,
new analytic bootstrap techniques are becoming applicable and should be vigorously pursued being potentially more general.
They appear especially advantageous for observables like four-point correlation functions,
which can now be studied to unprecedented precision.

In this paper, we study the four-point function of general half-BPS states in
maximally supersymmetric super Yang-Mills theory, or $\mathcal{N}=4$ SYM for short,
dual to type IIB superstring theory on AdS${}_5\times$S${}_5$ geometry.
Through the AdS/CFT correspondence, half-BPS operators are dual to $S_5$ spherical harmonics (Kaluza-Klein modes) of the graviton,
and we will consider the general correlator of four such spherical harmonics.
This provides an essential ingredient for studying the duality at loop level, where all modes can run inside the loop,
and also provides an explicit model for how local physics in an internal manifold gets encoded from the CFT perspective.

Four-point correlators have been much studied since the early days of the AdS/CFT correspondence.
Initially, techniques were developed for computing tree-level Witten diagrams in position space,
which led to the first complete results for the correlators of four dilatons or stress tensor multiplets \cite{DHoker:1999kzh,Arutyunov:2000py}.
In the case of the SYM model, part of these correlators are fixed by non-renormalization theorem and agree
with calculations at weak `t Hooft coupling \cite{Lee:1998bxa,Eden:2000bk},
but there is in addition an unprotected part which contains interesting dynamical information.
For example, a suitable limit reveals the flat space S-matrix of the underlying ten-dimensional
superstring theory \cite{Gary:2009ae,Penedones:2010ue,Maldacena:2015iua}.
This program was gradually extended to higher
spherical harmonics \cite{Arutyunov:2002fh,Arutyunov:2003ae,Berdichevsky:2007xd,Uruchurtu:2008kp,Uruchurtu:2011wh}.

Meanwhile, by analyzing the operator product expansion (OPE), it was discovered
how to exploit the large gap property of holographic theories, that is
the fact that all non-protected single-traces get lifted from the spectrum.
This led to various conjectures for correlators \cite{Nirschl:2004pa,Dolan:2006ec}.
However, it took almost a decade for a conjecture to appear in the general case \cite{Rastelli:2016nze},
thanks to a remarkably simple pattern which was observed in Mellin-space \cite{Penedones:2010ue,Fitzpatrick:2011ia}.
The corresponding position space calculation of tree-level Witten diagrams was only
very recently completed after a longstanding effort \cite{Arutyunov:2017dti,Arutyunov:2018tvn},
which confirmed the conjecture in all considered examples.

In this paper we will revisit correlators of half-BPS operators from the perspective of the analytic bootstrap,
which offers a constructive method to build correlators from their light-cone singularities \cite{Fitzpatrick:2012yx,Komargodski:2012ek,Alday:2015ewa}.
This can be formulated nonperturbatively as a Kramers-Kronig type dispersion relation,
known as the Lorentzian inversion formula, which explicitly reconstructs OPE data from a suitably defined absorptive part
\cite{Caron-Huot:2017vep, Simmons-Duffin:2017nub, Kravchuk:2018htv}.
For holographic theories at tree-level, the absorptive part is a finite sum of conformal blocks, thus reducing the computation
in a given theory to essentially a group-theoretic exercise (see also \cite{Alday:2017gde,Liu:2018jhs}).
The result is unique up to AdS contact interactions, whose size can be bounded by considering the Regge limit \cite{Alday:2016htq,Caron-Huot:2017vep}. These are further restricted by supersymmetry and in fact absent in $\mathcal{N}=4$ SYM as we will see.
The method, therefore, rigorously and unambiguously determines the correlators.
Interestingly,  in analogy with the unitarity method for one-loop S-matrices \cite{Bern:1994cg},
``squaring'' this tree-level four-point data gives sufficient information to go to one-loop \cite{Aharony:2016dwx}, 
as has now been successfully carried out for low spherical harmonics \cite{Aprile:2017bgs,Alday:2017xua,Alday:2017vkk}.

Study of $\mathcal{N}=4$ SYM specifically is further motivated by the possibility of finding unexpected structures.
In the planar limit, this theory is controlled by a rich and remarkable integrable system, making it a rather unique nontrivial
four-dimensional quantum field theory in which the
spectrum can be computed exactly at finite `t Hooft coupling, see \cite{Beisert:2010jr,Gromov:2014caa}.
It is important to determine what, if anything, of this structure persists beyond the planar limit.
Correlation functions at strong coupling offer a clean, natural environment in which to discuss this question.
Furthermore, we expect the correlators in this limit to provide important cross-checks and guidance as
integrability-based methods are now being developed to tackle the $1/N$ expansion \cite{Bargheer:2017nne,Ben-Israel:2018ckc},
as well as novel AdS-based techniques \cite{Cardona:2017tsw,Yuan:2017vgp,Yuan:2018qva}.

Hints of a hidden symmetry in correlators have been uncovered recently \cite{Aprile:2018efk},
who studied the anomalous dimension matrix governing the mixing between double-trace operators built from different S${}_5$ spherical harmonics, starting from the conjectured Mellin-space formula of \cite{Rastelli:2016nze}.
Amazingly, the eigenvalues of the anomalous dimension matrix are simple rational numbers,
for which the authors of ref.~\cite{Aprile:2018efk} conjectured a general formula.


This paper is organized into two parts.
First in section \ref{Sec:Generalities} we review the three ideas behind our calculations: the Operator Product Expansion,
the supersymmetric Ward identities satisfied by $\mathcal{N}=4$ SYM correlators, and the Lorentzian inversion formula.
As a warm-up, we then apply these formulas at order $1/N_c^0$ in section \ref{Sec:1/c0}.
Including the identity in two cross-channels yields double-trace data in the direct channel.
While conceptually straightforward, the supersymmetric OPE differs from the usual generalized free field result in simple but important ways.
Section \ref{Sec:Tree-level} contains our main computations: the tree-level correlator (order $1/N_c^2$) at
strong `t Hooft coupling. The full result is determined by just the singular part arising from half-BPS single-trace operators.
While their couplings are well-known and could be obtained from non-renormalization theorems \cite{Lee:1998bxa},
we show that crossing symmetry in fact suffices to bootstrap them from scratch.
This computation yields double trace mixing matrices, and we confirm the conjecture of \cite{Aprile:2018efk} in many examples.

In a second part of this paper, section \ref{sec:10D}, we explore the remarkable structure
found in this result. We will attribute the simplicity of the eigenvalues to an unexpected SO(10,2) symmetry,
representing an effective conformal invariance of ten-dimensional supergravity at tree-level.
This symmetry will be used to unify all spherical harmonics into a single ten-dimensional object.
Proof of the symmetry at the level of four-point correlators amounts to showing that this generating function
correctly predicts the singular part of each correlator; we check this for a large number of examples.
We also show how the symmetry readily implies the conjectured Mellin-space formula of \cite{Rastelli:2016nze},
and use it to obtain the leading-logarithmic term at each order in $1/N$.
Implications are discussed in the conclusion.

\section{Generalities} \label{Sec:Generalities}

At any coupling, $\mathcal{N}=4$ SYM contains a special class of operators,
which are half-BPS and thus annihilated by the maximal possible number of supercharges.
They transform as Lorentz scalars and
traceless symmetric tensors with respect to the SO(6)${}_R\simeq$ SU(4)${}_R$
global symmetry (Dynkin label $\big[0,p,0\big]$).
In the Lagrangian description of $\mathcal{N}=4$ SYM,
they are described as symmetrical traceless polynomials in the 6 adjoint scalar fields $\phi^a$ of the theory.
Denoting $x^\mu$ the spacetime coordinates and introducing null six-vectors $y^a$ to parametrize the SO(6)${}_R$ dependence,
they can be written as
\be \label{def Op}
 \OO^p(x,y) \propto {\rm Tr}\big[ (\phi^a y^a)^p\big] + O(1/N^2) \mbox{ corrections},
\ee
which will be normalized in this paper so that
\be
\big\l \OO^p(x_1,y_1)\OO^p(x_2,y_2)\big\r = \left( \frac{y_{12}^2}{x_{12}^2}\right)^p
\ee
where $x_{ij}^2=(x_i-x_j)^2$, $y_{ij}^2=y_i\cdot y_j$.
Due to the BPS condition, the scaling dimension of $\OO^p$ is exactly $p$, where $p\geq 2$.
At strong `t Hooft coupling, they admit a dual description as
Kaluza-Klein modes of the graviton on AdS${}_5\times$ S${}_5$.
According to the gravity analysis, all other operators in this limit are either multi-trace composites
built of products of the $\OO^p$, or are heavier than the string scale, $\Delta\gsim \lambda^{1/4}$.

Four-point correlators depend on AdS${}_5$ and S${}_5$ cross-ratios:
\begin{align}
u= \frac{x_{12}^{2}x_{34}^2}{x_{13}^2x_{24}^2} = z \bar{z}, \qquad &v = \frac{x_{23}^2x_{14}^2}{x_{13}^2x_{24}^2} = (1-z)(1-\bar{z}), \label{cross-ratios u,v} \\
\sigma= \frac{y_{12}^2 y_{34}^2}{y_{13}^2 y_{24}^2} = \alpha \bar{\alpha}, \qquad& \tau = \frac{y_{23}^2 y_{14}^2}{y_{13}^2 y_{24}^2} = (1-\alpha)(1-\bar{\alpha}), \label{cross-ratios sigma tau}
\end{align}
up to an overall prefactor conventionally written as follows \cite{Dolan:2003hv}:
\begin{align}
\big\l \OO^{p_1}(x_1,y_1) \cdots \OO^{p_4}(x_4,y_4) \big\r
\equiv& 
\left(\frac{y_{12}^2}{x_{12}^2}\right)^{\frac{p_1+p_2}{2}}
\left(\frac{y_{34}^2}{x_{34}^2}\right)^{\frac{p_3+p_4}{2}}
\left(\frac{x_{14}^2 y_{24}^2}{x_{24}^2 y_{14}^2}\right)^{\frac{p_{2}-p_1}{2}}
\left(\frac{x_{14}^2 y_{13}^2}{x_{13}^2 y_{14}^2}\right)^{\frac{p_{3}-p_4}{2}}
\nl
&\times \mathcal{G}_{\{p_i\} }(z,\zb,\a,\ab)\,.
\label{Eq:OriginalCorrelator}
\end{align}

We will be studying the correlator in the regime appropriate to the gauge-gravity duality,
where the 't Hooft coupling $\lambda = g_{YM}^2 N$ is large but finite, and work order by order in the
gravity loop expansion $1/c$ where $c=\frac{N^2-1}{4}$:
\begin{equation}
\mathcal{G}_{\{p_i\} } = \mathcal{G}_{\{p_i\}}^{(0)} + \frac{1}{c}\mathcal{G}_{\{p_i\}}^{(1)} + \frac{1}{c^2}\mathcal{G}_{\{p_i\}}^{(2)} + ...
\label{Eq:CorrelatorExpansion}
\end{equation}

The Operator Product Expansion (OPE) offers a series expansion of the correlator around kinematic limits.
In its most straightforward form, where we do not try to exploit supersymmetry,
it involves standard four-dimensional conformal blocks, times S${}_5$ spherical harmonics for each
$R$-symmetry representation which can be exchanged.
The conformal blocks for a given spin and dimension are written as
\begin{align}
G_{\ell,\Delta}^{r,s} (z,\bar{z}) &= \frac{z \bar{z}}{\bar{z}-z} \left[ k_{\frac{\Delta-\ell-2}{2}}^{r,s}(z) k_{\frac{\Delta+\ell}{2}}^{r,s}(\bar{z}) - k_{\frac{\Delta+\ell}{2}}^{r,s}(z) k_{\frac{\Delta-\ell-2}{2}}^{r,s}(\bar{z}) \right], \label{Eq:G-block}
\\
k_h^{r,s}(z) &= z^h \; _{2}F_1 \left(h+ \frac{r}{2}, h + \frac{s}{2}; 2h, z \right),
\end{align}
with $r=p_{21}\equiv p_2-p_1$, $s=p_{34}$. Since SO(6)${}_R$ and the SO(4,2) Lie algebra are analytic continuations of each other,
the S${}_5$ spherical harmonics admit identical expressions up to reversal of some quantum numbers:
\be
Z_{m,n}^{r,s} (\a,\ab) = (-1)^{m} G_{m,-n}^{-r,-s}(\a,\ab),
\ee
where the labels $m,n$ are related to the R-symmetry Dynkin labels by $\big[m, n-m, m\big]$.
In the literature these are often written as Jacobi polynomials, see \cite{Bissi:2015qoa}.
The OPE decomposition of the correlator (\ref{Eq:OriginalCorrelator}), in the $(12)$ or $s$-channel,
is then
\be
 \mathcal{G}_{\{p_i\} }(z,\zb,\a,\ab) = \sum_{\ell,\Delta,m,n}
\tilde{c}_{\{p_i\}}(\ell,\Delta,m,n)
 G_{\ell,\Delta}^{p_{21},p_{34}}(z,\zb)Z_{m,n}^{p_{21},p_{34}}(\a,\ab). \label{naive OPE}
\ee
Here the spin $\ell\geq 0$ is an arbitrary nonnegative integer and $\Delta$ runs over all operators with the given spin and $R$-symmetry.
The summations over $m,n$ however are finite, ranging over the R-symmetry representations which can appear in
the tensor product of each of the pairs $(1,2)$ and $(3,4)$.
Using the general formula for the tensor product of SU(4) representations (see \cite{Dolan:2002zh, Bissi:2015qoa}):
\begin{equation}
\big[0,p,0 \big] \times \big[0,q,0 \big] = \sum_{m=0}^{p} \sum_{s=0}^{p-m} \big[ m, q-p +2s, m\big],
\label{Eq:TensorProductStates}
\end{equation}
where we have assumed $p \leq q$,
we get the summation range in eq.~(\ref{naive OPE}):
\be
 0\leq m\leq {\rm min}\{ p_i\}, \qquad {\rm max}(|p_{12}|, |p_{34}|)+m \leq n \leq {\rm min}( p_1+p_2, p_3+p_4)-m,
\ee
where the difference between $n$ and its lower/upper bound is restricted to be an even integer.

The OPE~(\ref{naive OPE}) accounts for all the bosonic symmetries of the correlator. It is
still rather redundant because it does not exploit supersymmetry.
A natural refinement would be to use superconformal blocks instead, but here we will follow a simpler
route which is applicable thanks to the half-BPS nature of our external operators.


\subsection{Superconformal Ward identities}

A half-BPS supermultiplet is annihilated
by half of the 32 supercharges of the theory.  Since the remaining charges split into pairs of raising/lowering operators
acting within the multiplet, only 1/4 of the supercharges actually act nontrivially on a given bosonic primary $\OO^p(x,y)$.
This is significant because for four external operators, these 1/4 are generically linearly independent
and span the full algebra.  Thus the correlators of superconformal descendents
are fully determined from those of the primaries \cite{Chicherin:2014uca,Bissi:2015qoa}.

Linear independence however fails when the $x$ and $y$ cross-ratios are aligned in a specific way;
this leads to a Ward identity satisfied by the bosonic correlator \cite{Nirschl:2004pa, Dolan:2004mu, Bissi:2015qoa}:
\be
 \partial_z \left(\mathcal{G}(z,\zb,\a,\ab)\Big|_{\a=z}\right) =0. \label{ward identity}
\ee
That is, the $z$ dependence of the correlator disappears upon setting $\a=z$.

Since the dependence of the correlator on $\a,\ab$ is purely rational
(the left-hand-side of eq.~(\ref{Eq:OriginalCorrelator}) being polynomial in the $y_{ij}^2$), the Ward identities can be solved
by factoring out powers of $z-\a$, together with its conjugates under the
$(z\leftrightarrow \zb)$ and $(a\leftrightarrow\ab)$ symmetries. The most general solution, consistent
with these symmetries, is \cite{Dolan:2004iy,Bissi:2015qoa}:
\begin{align}
\mathcal{G}_{\{p_i\} }(z, \zb, \a, \ab) &= k \chi(z, \a) \chi(\zb, \ab) + \frac{(z-\a)(z-\ab)(\zb-\a) (\zb - \ab)  }{(\a - \ab)(z - \zb)} \nonumber \\
&\times \left(- \frac{ \chi(\zb,\ab) f(z,\a)}{\a z (\zb-\ab)} + \frac{\chi(\zb,\a) f(z,\ab)}{\ab z (\zb-\a)} + \frac{\chi(z,\ab) f(\zb,\a)}{\a \zb (z-\ab)} - \frac{\chi(z,\a) f(\zb,\ab)}{\ab \zb (z-\a)} \right) \nonumber \\
&+ \frac{(z-\a) (z - \ab) (\zb- \a) (\zb - \ab)}{(z \zb)^2 (\a \ab)^2} \HH_{\{p_i\}}(z,\zb, \a, \ab), \label{G ansatz}
\end{align}
where $\chi$ is a fixed function satisfying $\chi(z,z)=1$, given shortly.
Note that all the functions above depend on $\{p_i\}$, which we omitted in the above for readability.
In practice, starting from a correlator which fulfills the Ward
identity (\ref{ward identity}), $k_{\{p_i\} }$, which we will call the \textit{unit} contribution, is obtained simply
by setting $z=\a$, $\zb=\ab$. The \textit{chiral correlator}
$f_{\{p_i\} }$ is obtained by taking only one such limit and subtracting the unit:
\be\label{kf from G} \begin{aligned}
 k_{\{p_i\} } &= \GG_{\{p_i\} }(z, \zb, z, \zb), \\
 f_{\{p_i\} }(\zb,\ab) &= \frac{\ab \zb}{\zb - \ab} \left( \GG_{\{p_i\} }(z, \zb, z, \ab) - k_{\{p_i\}}\chi_{\{p_i\} }(\zb,\ab)  \right).
\end{aligned}\ee
Finally, the \textit{reduced correlator} $\HH_{\{p_i\} }$ can be extracted from $\GG_{\{p_i\} }$ by subtracting everything else that comes before it
in eq.~(\ref{G ansatz}).

There is a rather unique, convenient choice for the function $\chi(z,\ab)$, which ensures that the superconformal Casimir
equation commutes with the preceding decomposition \cite{Bissi:2015qoa}:
\begin{equation}
\chi_{\{p_i\}}(z,\a) = \left( \frac{z}{\alpha} \right)^{\max(|p_{21}|,|p_{34}|)/2} \left( \frac{1-a}{1-z} \right)^{\max(p_{21}+p_{34},0)/2}.
\label{Eq:ChiWard}
\end{equation}
The Casimir operator then annihilates the $k$ contribution, in particular.
In fact there are four possible solutions to this constraint, obtained by analytic continuation in the $p_i$'s:
the above solution is singled out by the fact that it does not introduce spurious negative exponents at $z\to 0$ and $\a\to 1$.
The same solution was used in \cite{Bissi:2015qoa} (who restricted to a specific ordering of the $p_i$'s).

We now review the implications of the Casimir equations, following \cite{Bissi:2015qoa}.
Its action on the \textit{chiral correlator} takes on a separated form, 
whose general solution involves products of the hypergeometric functions in eq.~(\ref{Eq:G-block}), thus giving the OPE:
\begin{equation}
f_{\{p_i\}}(z,\a) =  \sum_{j=0}^\infty \sum_m b_{\{p_i\}}(j,m) k_{1+m/2+j}^{p_{21},p_{34}}(z) k_{-m/2}^{-p_{21},-p_{34}}(\a).
\label{Eq:fGeneric}
\end{equation}
The sum over $m$ is finite since $f$ is a polynomial in $1/\a$, whose degree determines the range:
\be
 \text{max}(|p_{12}|,|p_{34}|)\  \leq\  m\  \leq\  \text{min}(p_1+p_2,p_3+p_4) -2,
\ee 
where in addition $m$ should differ from its lower bound by an even integer.
Single-valuedness of the correlator (\ref{Eq:OriginalCorrelator}) forces $j$ to be an integer.
It must be nonnegative due to the unitarity bound, since the superconformal
Casimir eigenvalue $(m+j)(m+j+1) - m(m+1)$ must be nonnegative.

The Casimir equation for the \textit{reduced correlator} $\HH$ also takes separated form,
and its solution is similar to the naive OPE (\ref{naive OPE}) but with the dimension shifted by $4$
in accordance to the $z\zb$ denominator in the prefactor multiplying $H$:
\begin{equation}
\HH_{\{p_i\}}(z,\zb,\a,\ab)=\sum_{\ell,\Delta,m,n} a_{\{p_i\}}(\ell,\Delta,m,n)
G_{\ell,\Delta+4}^{p_{21},p_{34}} (z,\zb) Z_{m,n}^{p_{21},p_{34}}(\a,\ab).
\label{Eq:Hpart}
\end{equation}
The representations appearing in $H$ can be worked out from the $\a$, $\ab$ dependence of the coefficient of $\HH$ in eq.~(\ref{G ansatz}),
and are only those which appear in the tensor products of $[0,p_i-2,0]$; this gives the range
\be
 0\leq m\leq {\rm min}\{ p_i-2\}, \qquad {\rm max}(|p_{12}|, |p_{34}|)+m\ \leq\ n\ \leq\ {\rm min}(p_1+p_2, p_3+p_4)-m-4.
\label{small range}
\ee
While formally similar to the bosonic expansion enjoyed by $\mathcal{G}$ in eq.~(\ref{naive OPE}), 
we see that the OPE for the susy-decomposed correlators $f$ and $\HH$ contains much fewer terms.
For example, for the simplest case of the
2222 correlator, eq.~(\ref{naive OPE}) contains six distinct R-symmetry representations
whereas there is a single one in eq.~(\ref{Eq:Hpart}),
involving only $Z_{0,0}^{(0,0)}(\a,\ab)=1$.  The full $\a$ dependence is then entirely produced
by the chiral correlator $f(z,\a)$ and the various factors in (\ref{G ansatz}).

\subsection{Lorentzian inversion formula}
\label{Sec:InversionIntegral}

To determine the OPE data $b^{\{p_i\}}_{j,m}$ and $a^{\{p_i\}}_{\ell,\Delta,m,n}$
entering eqs.~(\ref{Eq:fGeneric}) and eqs.~(\ref{Eq:Hpart}),
our main tool will be a recently derived inversion integral which plucks out
the OPE coefficients from the correlator.
The key point is that this formula does not require to know the full correlator,
but just its ``double discontinuity'' which is much easier to compute.

Let us first deal with the global symmetry.
Using that the $Z_{m,n}(\a,\ab)$ represent S${}_5$ spherical harmonics, which are mutually orthogonal,
it is straightforward to write down an integral which projects the reduced correlator
onto any desired representation.  As noted, the $Z_{m,n}$'s are secretly Jacobi polynomials,
with argument $\cos\theta=1-\frac{2}{\a}$, which suggests that the natural integration range is $\cos\theta\in [-1,1]$
or $\a,\ab\in [1,\infty)$.
One can then show that the following orthogonality relation holds:
\be
 \HH_{\{p_i\}}(z,\zb,m,n)
 = \left(8\pi^4 \tilde\kappa^{r,s}_{2+\frac{m+n}{2}}\tilde\kappa^{r,s}_{1+\frac{n-m}{2}}\right)^{-1}
 \int_1^\infty \frac{d\a}{\a^2}\frac{d\ab}{\ab^2} \left( \frac{\a-\ab}{\a \ab}\right)^2
Z^{-r,-s}_{m,n}(\a,\ab) \HH_{\{p_i\}}(z,\zb,\a,\ab)
\label{Eq:Z projection}
\ee
where $r=p_{21}$, $s=p_{34}$,  and
\be
\tilde{\kappa}^{r,s}_h = \frac{\Gamma(h+\frac12r)\Gamma(h-\frac12r) \Gamma(h+\frac12s)\Gamma(h-\frac12s)}{2\pi^2 \Gamma(2h-1) \Gamma(2h)} = \frac{r_{h}^{r,s}\, r_h^{-r,-s}}{2\pi^2(2h-1)},
\ee
\be 
r_h^{r,s} = \frac{\Gamma(h+\frac{1}{2}r) \Gamma(h+\frac{1}{2}s)}{\Gamma(2h-1)}.
\label{kappa tilde}
\ee
This diagonalizes the S${}_5$ part of the OPE (\ref{Eq:Hpart}), leaving only sums over dimension and spin:
\label{rfct}
\be
 \HH_{\{p_i\}}(z,\zb,m,n) = \sum_{\ell,\Delta} a_{\{p_i\}}(\ell,\Delta,m,n) G_{\ell,\Delta+4}^{(p_{21},p_{34})} (z,\zb).
 \label{OPE Hmn}
\ee
The Lorentzian inversion formula inverts this $z,\zb$ OPE:
it extracts the coefficient of a given $G_{\ell,\Delta}$ block by integrating against $z,\zb$.
Writing its left-hand-side as $c_{m,n}(\ell,\Delta-4)$
to compensate for the argument of the block being $\Delta+4$, the formula from \cite{Caron-Huot:2017vep} is written as a sum of $t$-channel and
$u$-channel contributions:
\be\begin{aligned}
c_{\{p_i\}}(\ell,\Delta-4,m,n) &= \frac14\tilde{\kappa}^{p_{21},p_{34}}_{\frac{\Delta+\ell}{2}+2}
\displaystyle \int_0^1 \frac{dz}{z^2} \frac{d\zb}{\zb^2} \left( \frac{\bar{z}-z}{\bar{z}z} \right)^2
G_{\Delta-3,\ell+3}^{-p_{21},-p_{34}}(z,\zb)
\dDisc\big[\HH_{\{p_i\}}(z,\zb,m,n)\big]
\\ &+(-1)^{\ell+m} \big( p_1\leftrightarrow p_2\big)
\label{Eq:cInversion}
\end{aligned}\ee
with $p_{ij}=p_i-p_j$.\footnote{
Comparison with \cite{Caron-Huot:2017vep} requires the identity:
$\big[(1-z)(1-\zb)\big]^{r+s}
G_{\Delta-3,\ell+3}^{r,s}(z,\zb) = G_{\Delta-3,\ell+3}^{-r,-s}(z,\zb)
$.
}
The function $c_{\{p_i\}}(\ell,\Delta,m,n)$ serves as a generating function of the OPE data, which, for a fixed integer $\ell$,
are encoded in its $\Delta$-plane poles:
\begin{equation}
\lim_{\Delta\to \Delta_k} c_{\{p_i\}}(\ell,\Delta,m,n) = \displaystyle \frac{a_{\{p_i\}}(\ell,\Delta_k,m,n)}{\Delta_k - \Delta},
\label{naive c poles}
\end{equation}
where $k$ labels the the different exchanged operators. 

The double discontinuity in the integrand is a natural Lorentzian object
defined as the expectation value of a double commutator $\frac12\l 0| [\OO_4,\OO_1][\OO_2,\OO_3]|0\r$ in
a region where the two invariants $x_{14}^2$ and $x_{23}^2$ are timelike \cite{Caron-Huot:2017vep}.
Since $(1-\zb) \propto x_{14}^2x_{23}^2$ switches sign when either invariant crosses the light-cone,
the different operator orderings can be reached by analytically continuing around $\zb=1$ with $z$ fixed.
Specifically, working in the region $0<z,\zb<1$ and denoting
the usual Euclidean-branch correlator simply as $\mathcal{G}$, the double commutator is equal to:
\be
\dDisc \big[\mathcal{G}(z,\bar{z})\big] = \cos(\pi\alpha) \mathcal{G}(z,\bar{z}) -
\tfrac{1}{2}e^{i\pi\alpha} \mathcal{G}^{\Circlearrowleft}(z, \bar{z}) - \tfrac{1}{2} e^{-i\pi\alpha}\mathcal{G}^{\Circlearrowright}(z, \bar{z}),
\label{Eq:doubleDisc}
\ee
where $\alpha= \frac{p_{21}+p_{34}}{2}$.

It is important to stress that eq.~(\ref{Eq:cInversion}), contrary to eq.~(\ref{Eq:Z projection}),
is not a straightforward consequence of an orthogonality condition applied termwise to the OPE.
In fact, the {\rm dDisc} operation naively annihilates each term in eq.~(\ref{OPE Hmn}).
In reality it is nonzero because the OPE sum diverges for $\zb>1$ which generally creates a branch cut or pole at $\zb=1$.
The formula is thus more analogous
to a Kramers-Kronig dispersion relation which reconstructs a function from its discontinuities, or the Froissart-Gribov
formula in the scattering amplitude context.

The OPE data in one channel is thus obtained from that
in two cross-channels, which can be used to efficiently compute the two $H$ functions on the right of eq.~(\ref{Eq:cInversion}).
This reorganization of crossing symmetry is advantageous in large-$N$ theories,
because the $\dDisc$ operation kills cross-channel operators with double-trace dimensions, see \cite{Caron-Huot:2017vep}.
Only single-traces contribute in the cross-channels
(up to order $1/c^2$, that is one-loop order in supergravity).   We will see this in action in the next sections.

Using the explicit form (\ref{Eq:G-block}) for the conformal blocks,
the formula (\ref{Eq:cInversion}) can be written in a more useful factorized form as
\be\begin{aligned}
c_{\{p_i\}}(\ell,\Delta,m,n) &= \frac{\tilde{\kappa}_\hb^{p_{21},p_{34}}}{2}\!\!
\displaystyle \int_0^1 \frac{dz}{z^2} \frac{d\zb}{\zb^2}
\ k_{1-h}^{-p_{21},-p_{34}}(z)k_{\hb}^{-p_{21},-p_{34}}(\zb)
\dDisc\left[\frac{\zb-z}{\zb z}\HH_{\{p_i\}}(z,\zb,m,n)\right]
\\ & \hspace{5mm} +
(-1)^{\ell+m} \big( p_1\leftrightarrow p_2\big), 
\label{Eq:cInversion 2}
\end{aligned}\ee
where $h=1+\frac{\Delta-\ell}{2}$ and $\hb=2+\frac{\Delta+\ell}{2}$. In practice, the OPE data is encoded in poles with respect to $h$ which comes from the $z\rightarrow 0$ limit of the $z$ integration. The $\bar{z}$ integral is dual to $\bar{h}$.
Analytic results for the integrals we will need are given in appendix \ref{Appendix: Inversion}.

Some words about convergence.  In an abstract (unitary) CFT, convergence of the integral for $\ell>1$ (along the principal series $\Delta=d/2+i\nu$) was proven in \cite{Caron-Huot:2017vep} using boundedness of $\mathcal{G}$ from the Regge limit $z,\zb\to 0$ (more precisely,
boundedness of each term
in eq.~(\ref{Eq:doubleDisc})).  For the present supersymmetric correlator, tracing through the factors in eq.~(\ref{G ansatz}),
boundedness of $\mathcal{G}$ actually implies that $H/(z\zb)^2$ is bounded. Supersymmetry thus buys
us four units of spin: convergence is ensured for $\ell >-3$, that is, all spins.\footnote{A simply analogy is the flat space
supergravity $S$-matrix, where supermomentum conservation $\delta^{(16)}(Q)$ removes a factor $s^4$ in the Regge limit $s\to\infty$.}

Order by order in the $1/c$ expansion, the situation can be less favorable than nonperturbatively, reflecting the
worsening ultraviolet behavior of supergravity with increasing loop order.
At order $1/c$, convergence for $\ell>2$ was nonetheless proven in \cite{Caron-Huot:2017vep} from a stress tensor sum rule,
which thus ensures convergence for $\ell>-2$ in our supersymmetric case.
Therefore, at order $1/c$, the integral (\ref{Eq:cInversion 2}) is guaranteed to recover \emph{all} OPE data in $\mathcal{N}=4$ super Yang-Mills.
The coefficients $c(\ell,\Delta,m,n)$ contain information about both protected \emph{and} unprotected contributions,
which are effectively treated on the same footing.

Finally, for the chiral correlator $f$, the OPE (\ref{Eq:fGeneric}) can be similarly inverted
using the one-dimensional inversion integral from ref.~\cite{Simmons-Duffin:2017nub}.
This formula deals with the case where the sum runs over integer $h$, which is precisely the case for the chiral correlator!\footnote{
 We find that the result in \cite{Simmons-Duffin:2017nub} holds also in the case of non-identical external operators 
 provided that the sum runs over half-integer $h$ when $p_{12}$ is odd.}
Using orthogonality on a simple contour integral around $\a=0$ to deal with the $R$-symmetry part,
\be
 f_{\{p_i\}}(z,m) = \frac{1}{2\pi i}\oint \frac{d\a}{\a^2} k_{1+m/2}^{p_{21},p_{34}}(\a) f_{\{p_i\}}(z,\a),
\ee
the inversion integral then gives the coefficient as
\be
b_{\{p_i\}}(j,m)  = \int_0^1 \frac{d\zb}{\zb^2} k_{1+m/2+j}^{-p_{21},-p_{34}}(\zb) \ \dDisc\big[f_{\{p_i\}}(\zb,m)\big].
 \label{f Inversion}
\ee
In this way
all protected and unprotected data for $f$ and $H$ in eqs.~(\ref{Eq:fGeneric}) and \eqref{Eq:Hpart} are recovered from the double-discontinuity.

We will be particularly interested in double-trace operators $[\mathcal{O}_p \mathcal{P}_q]_{k,\ell}$,
which in the large-$N$ limit should appear as poles near twist $\Delta-\ell= p+q+2k+\gamma$, 
that is $h=1+k+\frac{p+q}{2}+\frac{\gamma}{2}$, where $\gamma$ is a small anomalous dimension.
Actually, many nearly-degenerate operators generally contribute, so summing over them, eq.~\eqref{naive c poles} becomes
\begin{equation}
c_{\{p_i\}}(h,\hb=h+\ell+1,m,n) = \left\langle \frac{a_{\{p_i\}}(\ell,\Delta,m,n)}{2(1+k+\frac12(p+q)+\frac{1}{2}\gamma - h)} \right\rangle,
\label{Eq:cGamma}
\end{equation}
where the brackets imply that we sum over all superconformal primary operators with spin $\ell$ and approximate twist $p+q+2k$,
and $R$-symmetry representation $[m,n-m,m]$.
One can use this equation to organize the OPE data as a $1/c$ expansion, with $c=\frac{N^2-1}{4}$.
Indeed, one can consider the $1/c$ expansion of the scaling dimension and OPE coefficients as
\begin{align}
\Delta_{n,\ell} &= \Delta + \frac{1}{c} \gamma_{n,\ell}^{(1)} + \frac{1}{c^2} \gamma_{n,\ell}^{(2)} +... \label{Eq:massExpansion} \\
a_{n,\ell} &= a_{n,\ell}^{(0)} + \frac{1}{c} a_{n,\ell}^{(1)} + \frac{1}{c^2} a_{n,\ell}^{(2)} + ...
\label{Eq:aExpansion}
\end{align}
In particular, expanding in $1/c$, leading anomalous dimensions $\gamma_{n,\ell}^{(1)}$ appear as double poles 
in the tree-level data $c_{\{p_i\}}^{(1)}$.

Below we will first recover the OPE data for $f$ and $H$ at order $1/c^0$ (disconnected correlator),
where only the identity contributes in the cross-channel, which we will resolve into protected and unprotected double-traces in the $s$-channel.
We will then proceed to order $1/c$, where we will only need to insert a finite number of single-trace half-BPS blocks in the cross-channels.

\subsection{Other decompositions in the literature}

Before proceeding, let us briefly contrast our organization scheme, based on the Ward identity (\ref{ward identity})
and Casimir equation, to other similar but distinct formulas used in the literature.
A common way of writing the OPE is to separate ``short'' (protected) and ``long'' (unprotected)
superconformal blocks, schematically:
\be
 \GG(z,\zb,\a,\ab) = \sum c^{\rm short} g^{\rm short}(z,\zb,\a,\ab) + \sum c^{\rm long} g^{\rm long}(z,\zb,\a,\ab).
\ee
In terms of the supersymmetry decomposition (\ref{G ansatz}), the short blocks contribute to all of $k,f,H$,
whereas long blocks contribute only to $H$ (as the single term in eq.~(\ref{Eq:Hpart})).
Another popular scheme exploits nonrenormalization theorems to separate
out contributions which do not depend on the coupling:
\be
 \GG(z,\zb,\a,\ab) = \GG^{\rm free}(z,\zb,\a,\ab) +
 \frac{(z-\a) (z - \ab) (\zb- \a) (\zb - \ab)}{(z \zb)^2 (\a \ab)^2} \HH^{\rm interacting}_{\{p_i\}}(z,\zb, \a, \ab)
\ee
where $\GG^{\rm free}$ can be computed from the limit of weak `t Hooft coupling $\lambda\to 0$
and $\HH^{\rm interacting}(z,\zb,\a,\ab)$ only receives contribution from long blocks (of both the strong and weakly interacting theories).
Since the $g^{\rm long}$ only contribute to the $H$ part, these schemes both isolate
the nontrivial corrections into $H$-like functions, which differ only by the amount of protected physics subtracted from it.

The decomposition used in this paper is distinct since we do not subtract anything from $H$,
neither its short block contribution nor its free theory limit.
Rather we directly obtain the functions $f$ and $H$ at strong coupling,
by computing their double-discontinuity and using the inversion integrals (\ref{Eq:cInversion 2}) and (\ref{f Inversion})
to find the expansion coefficients (\ref{Eq:fGeneric}) and (\ref{Eq:Hpart}).
This method eschews the use of nonrenormalization theorems and detailed knowledge of superconformal blocks.

\section{Disconnected correlator: leading order $1/c^{0}$} \label{Sec:1/c0}

\begin{figure}[h]
\centering
\def\svgwidth{\linewidth}
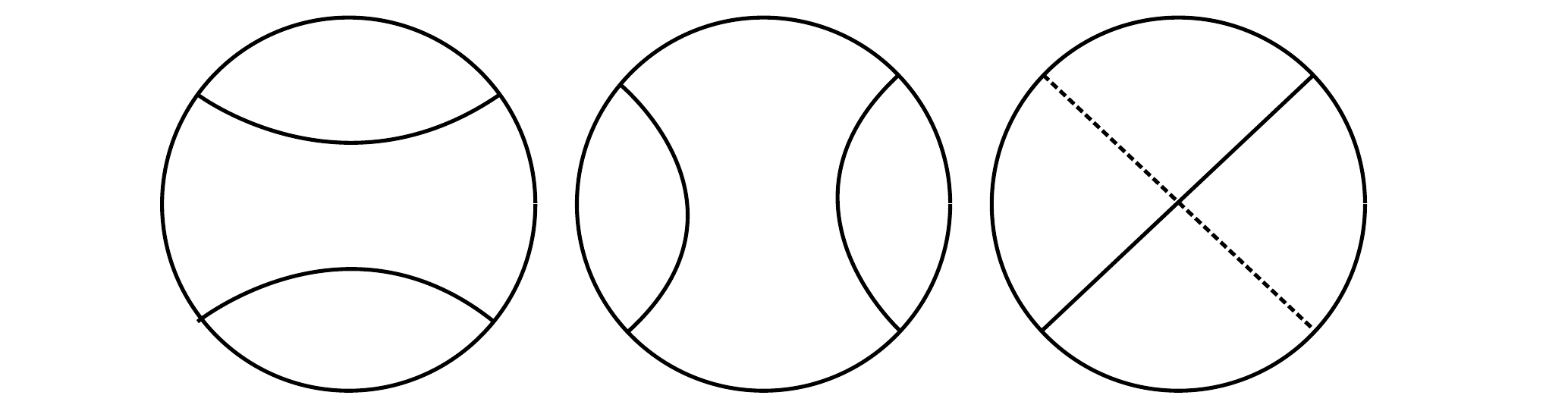
\caption{Possible Witten diagrams for the disconnected correlator at order $1/c^0$. When $p\neq q$, only the first diagram contributes. For $p=q$, the two other diagrams can appear. See eq.~\eqref{G disconnected}.}
\label{Fig:disconnected correlator}
\end{figure}

As a first step toward anomalous dimensions of double-trace operators, here we detail
the disconnected correlator OPE coefficients $\langle a^{(0)} \rangle_{pqqp}$, which set the overall normalization.
The disconnected correlator itself is simply (see fig.~\ref{Fig:disconnected correlator})
\be
 \mathcal{G}_{pqqp}^{(0)} =\delta_{p,q} +  \left(\frac{u}{\sigma}\right)^{\frac{p+q}{2}} \left[
 \left( \frac{\tau}{v}\right)^q + \delta_{p,q}\right]\,.
 \label{G disconnected}
\ee
The OPE decomposition of such correlators is well known \cite{Fitzpatrick:2011dm}.
However, we want to first apply the supersymmetry decomposition (\ref{G ansatz}).
While we find that the unit part is given by a simple general formula: $k^{(0)}_{pqqp} = 1+2\delta_{pq}$,
the chiral and reduced correlators are generally given by more lengthy expressions.
To give a few examples:
\be\begin{aligned}
 f^{(0)}_{2222} &= \frac{z^2(1-a)}{(1-z)^2a}+ \frac{z}{1-z} +z + \frac{z^2}{a^2},
 &\qquad H^{(0)}_{2222} &= u^2 +\frac{u^2}{v^2}\,,
\\
f^{(0)}_{2332}&= \frac{(1-a)z^{3/2}}{(1-z)^3 a^{3/2}}(a+z-2az),
 &\qquad H^{(0)}_{2332}&= \frac{\tau u^{5/2}}{\sigma^{1/2}v^3}.
\end{aligned}\ee
Generally, starting from $p,q\geq 3$, neither $f$ nor $H$ are simple sums and the expressions quickly grow lengthier with increasing $p$ and $q$.
Nonetheless, the corresponding OPE data admits a simple uniform expression, as we now describe.

\subsection{OPE coefficients} \label{Sec:a0}

The OPE coefficients $\langle a^{(0)}(h,\hb,m,n)\rangle_{pqqp}$ for a given correlator depend on $h,\bar{h}$, the scaling dimensions $p,q$ and the R-symmetry representation labels $m,n$ of the exchanged operator.
They are obtained by applying the R-symmetry projection (\ref{Eq:Z projection}) and inversion integral (\ref{Eq:cInversion 2}) to the double-discontinuity of $H^{(0)}$. (We will discuss the chiral part $f^{(0)}$ below.)
Since there are no branch cut at $\zb=1$, the double-discontinuity here simply picks out the polar terms.
For instance, in the above examples, switching to the variables $x=\frac{z}{1-z}$ and $y=\frac{1-\zb}{\zb}$, we find for the poles in $y$:
\ba
 \frac{\zb-z}{z\zb} \dDisc H_{2222}^{(0)} &=& Z_{0,0}^{0,0}(\alpha,\ab)\times \dDisc\left[\frac{x}{y^2}-\frac{x^2}{y}\right],\nonumber \\
 \frac{\zb-z}{z\zb} \dDisc H_{2332}^{(0)} &=& Z_{0,1}^{1,1}(\alpha,\ab)\times \frac{1}{\sqrt{z\zb}}\dDisc\left[\frac{x^2}{y^3}-\frac{x^3}{y^2}\right].
\ea
To give a less trivial example:
\be\begin{aligned}
 \frac{\zb-z}{z\zb} \dDisc H_{3333}^{(0)} &=
 Z_{0,2}^{0,0}(\a,\ab)\ \dDisc\left[\frac{x}{y^2}-\frac{x^2}{y}\right]
 + Z_{1,1}^{0,0}(\a,\ab)\ \dDisc\left[\frac{x^2}{2y^3}-\frac{x^3}{2y^2}-\frac{x}{y^3}+\frac{x^3}{y}\right]
\\
& + Z_{0,0}^{0,0}(\a,\ab)\ \dDisc\left[\frac{x^2}{6y^3}-\frac{x}{y^3}+\frac{x}{y^2} -(x{\leftrightarrow}y^{-1})\right].
\end{aligned}\ee
We notice that all expressions are antisymmetric under the interchange  $x\leftrightarrow y^{-1}$.
This stems from the manifest antisymmetry in $(z,\zb)$ of the left-hand-side together with the fact that the expressions are simple polynomials
in $x$ and $y^{-1}$ (after factoring the same power of $z$ and $\zb$ that appeared in the integral (\ref{Eq:cInversion 2})).
The integral can be immediately performed using the elementary integrals (\ref{Eq:zbarIntegral}) and (\ref{Eq:zIntegral}):
\begin{subequations}
\ba
\label{c2222 zero}
 c^{(0)}_{2222}(h,\hb,0,0)&=& r_h^{0,0}r_{\hb}^{0,0} \pi\cot(\pi h) \times  (1+(-1)^\ell) \Big(\hb(\hb-1)-h(h-1)\Big),\\
\label{c2332 zero}
 c^{(0)}_{2332}(h,\hb,0,1) &=& r_h^{1,1}r_{\hb}^{1,1} \pi\tan(\pi h) \times \tfrac12\Big(\hb(\hb-1)-h(h-1)\Big)(h-\tfrac12)(\hb-\tfrac12),
\ea
\end{subequations}
whereas in the $3333$ case the three non-vanishing R-symmetry structures give three coefficients:
$c^{(0)}_{3333}(h,\hb,0,2)$, $c^{(0)}_{3333}(h,\hb,1,1)$ and $c^{(0)}_{3333}(h,\hb,0,0)$.

Let us briefly interpret such results in terms of double-trace and protected operators.
Double-traces have twist $p+q+2k$ with $k=0,1,2\ldots$, which correspond to $h=\frac{p+q+2}{2}+k$
and therefore $h\geq 3$ in the $2222$ case.
However, we see from the formula (\ref{c2222 zero}) that $c_{2222}$ has poles (from the cotangent) at each integer $h$.
The poles at $h\leq 0$ can ignored (due to the terms discarded in doing the integral (\ref{Eq:zIntegral})),
but the poles at $h=1$ and $h=2$ would seem puzzling: these naively correspond to twists $0$ and $2$, which shouldn't be there.
Similarly, for $2332$, the formula exhibits two poles below the $h=\frac72$ double-trace threshold: at $h=\frac32$ and $\frac52$.
As we will see shortly, the resolution is that these
poles originate from short (and so-called semi-short) multiplets, which can contribute to the reduced correlator $H$
with the two apparent twists:
\be
 h= \frac{n-m+2}{2} \qquad\mbox{or}\qquad h= \frac{m+n+4}{2}\,. \label{protected h}
\ee 

We are now ready to state the OPE coefficients in the general case.
Due to the general form of the integrals (\ref{Eq:zbarIntegral})-(\ref{Eq:zIntegral}), the result is always
the $r$ factors times a polynomial in $h,\hb$.
This polynomial is easily determined using the following observations:
\begin{itemize}
\item It is odd under interchange of $h$ and $\hb$ (due to the $x\leftrightarrow y^{-1}$ symmetry just mentioned).
\item It is symmetrical under $\hb\to 1-\hb$ (up to an overall minus sign when $p+q$ is odd).
\item It vanishes for all values of $h$ below the double-trace threshold $h=\frac{p+q+2}{2}$,
\emph{except} at the protected locations allowed by eq.~(\ref{protected h}).
\end{itemize}
It is easy to show that the result is always simply the lowest-order polynomial with these properties,
up to a $h,\hb$-independent constant.  It can be written in general as:
\begin{align}
\langle a^{(0)}_{pqqp} \rangle(h,\hb,m,n) &=
\left(1+\delta_{p,q}(-1)^{\ell+m}\right)
\frac{r_{h}^{q-p,q-p}\,r_{\hb}^{q-p,q-p}}{r_{1+\frac{n-m}{2}}^{q-p,q-p}r_{2+\frac{m+n}{2}}^{q-p,q-p}} \nonumber
\\
&
\times
\frac{\Gamma(h+ \frac{p+q}{2}) \Gamma(\bar{h} + \frac{p+q}{2}) }{\Gamma(h-\frac{p-q}{2}) \Gamma(\bar{h}-\frac{p+q}{2})}
\frac{\hb(\hb-1)-h(h-1)}{\Delta_{m,n}^{(8)}(h,\bar{h})}
\nonumber \\
& \times \frac{(3+m+n)(1+n-m)(m+1)(n+2)}{\Gamma(\frac{p+q+4+m+n}{2}) \Gamma(\frac{p+q+2+n-m}{2}) \Gamma(\frac{p+q-(n-m)}{2}) \Gamma(\frac{p+q-(2+m+n)}{2})},
\label{Eq:a0pqqp}
\end{align}
where
\begin{align}
\Delta_{m,n}^{(8)}(h,\bar{h}) &= \left(h- \frac{n-m+2}{2} \right) \left(h+ \frac{n-m}{2} \right) \left( h - \frac{m+n+4}{2} \right) \left( h + \frac{m+n+2}{2} \right) \nonumber \\
&\times \left(\bar{h}- \frac{n-m+2}{2} \right) \left(\bar{h}+ \frac{n-m}{2} \right) \left(\bar{h} - \frac{m+n+4}{2} \right) \left(\bar{h} + \frac{m+n+2}{2} \right).
\label{Eq:Delta8}
\end{align}
Notice that the factor $\Delta^{(8)}$ simply provides poles
at the locations (\ref{protected h}) and their orbits under $h\mapsto 1-h$ and $h\mapsto \hb$.

For the chiral correlator $f$, which depends only on $z,\alpha$, we similarly apply the one-dimensional inversion formula
in eq.~(\ref{f Inversion}). We find a structurally similar result (loosely,
one drops all the factors which depend either on $h$ or $m+n$, and substitutes $\frac{n-m}{2}\mapsto \frac{m}{2}$):
\begin{align}
\langle b^{(0)}_{pqqp} \rangle (j,m)
&= \Big(1+\delta_{p,q}(-1)^{j}\Big) \frac{r_{\bar{h}}^{q-p,q-p}}{r_{1+m/2}^{q-p,q-p}}
\times \frac{\Gamma(\bar{h}+\frac{p+q}{2})}{\Gamma(\bar{h} - \frac{p+q}{2})}\frac{(-1)^{(p+q-m-2)/2}}{(\hb-m/2-1) (\hb+m/2)}\Bigg|_{\hb=1+m/2+j}
 \nonumber \\
&
\times \frac{m+1}{\Gamma(\frac{p+q+m}{2}+1)\Gamma(\frac{p+q-m}{2})}.
\label{Eq:a0chiral}
\end{align}
The last line of both eqs.~(\ref{Eq:a0pqqp}) and (\ref{Eq:a0chiral}) contains a $h,\hb$-independent normalization factor,
which in all cases is determined by the following nice property: the coefficient
becomes unity when setting $h=1+\frac{n-m}{2}$ and $\hb=2+\frac{m+n}{2}$ (or $j=0$ in the chiral case).
This is the expected half-BPS contribution to the correlator as we now explain.

\subsection{Reorganizing into superconformal blocks}

Although this will not be needed below, it is an interesting cross-check that the above results can be reorganized
into superconformal blocks.

A general expectation is that only operators with double-trace dimensions contribute to the disconnected correlator
(with the lone exception of the identity operator).  However, looking at the result
for $a^{(0)}_{pqqp}$ in eq.~(\ref{Eq:a0pqqp}), we found contributions to $H$ below the double-trace threshold $h\geq \frac{p+q+2}{2}$,
exemplified by eq.~(\ref{protected h}).  Looking more closely into them, we find that they always appear with identical coefficients
as contributions to the chiral part $f$ in eq.~(\ref{Eq:a0chiral}).
These together organize into superconformal blocks with double-trace dimensions.

\begin{table}[h]
\begin{center}
\begin{tabular}{||c|c|c||}
\hline
Multiplet & Dynkin labels & Dimension $\Delta$ and spin $\ell$ \\
\hline
Half-BPS $\mathcal{B}_{0,n}$& $\big[0,n,0\big]$& $\Delta=q$, $\ell=0$\\
Quarter-BPS $\mathcal{B}_{m,n}$& $\big[m,n{-}m,m\big]$& $\Delta = m+n$, $\ell=0$, $m\geq 1$ \\
Semi-short $\mathcal{C}_{\ell,m,n}$ & $\big[m,n{-}m,m\big]$& $\Delta=m+n+2+\ell$ \\
Long $\mathcal{A}_{\ell,\Delta,m,n}$ & $\big[m,n{-}m,m\big]$& $\Delta > m+n+2+\ell$ \\
\hline
\end{tabular}
\end{center}
\caption{Supermultiplets which can appear in the four point function of half-BPS operators.}
\label{Tab:Multiplets}
\end{table}

An effective way to group these contributions is to use the superconformal (quadratic) Casimir invariant.
The different types of supermultiplets that can contribute to the four-point function are listed in table \ref{Tab:Multiplets}
(see for e.g. \cite{Bissi:2015qoa}).
The corresponding Casimir eigenvalues are
\begin{align}
\label{C2 short}
\mathcal{C}_2 &\equiv \tfrac12\left(\Delta(\Delta+4)+\ell(\ell+2)-m(m+2)-n(n+4)\right)
\\
&= \left\{\begin{array}{l@{\hspace{10mm}}l}
0, & \mbox{Half-BPS } \mathcal{B}_{0,n}\,,\\
m(n+1), &\mbox{Quarter-BPS } \mathcal{B}_{m,n}\,,\\
(\ell+m+2)(\ell+n+3), & \mbox{Semi-short } \mathcal{C}_{\ell,m,n}\,. \end{array}\right.
\end{align}
As mentioned, the superconformal Casimir commutes with the
supersymmetry decomposition (\ref{G ansatz}) and is diagonalized by the OPE expansions (\ref{Eq:fGeneric}) and \eqref{Eq:Hpart}.
For convenience let us recall the eigenvalues:
\begin{align}
\label{C2 susy}
 \mathcal{C}_2 & \left\{ \begin{array}{l} k \\
 f_{j,m}(z,\a)\equiv k_{1+m/2+j}(z)k_{-m/2}(\a) \\
 H_{\ell,\Delta,m,n}(z,\zb,a,\ab)\equiv G_{\Delta+4,\ell}(z,\zb)Z_{m,n}(\a,\ab)
 \end{array}\right.
\quad \nonumber \\
=
& \ \left\{\begin{array}{l} 0 \\ j(m+j+1) \\ 
  \tfrac12\left(\Delta(\Delta+4)+\ell(\ell+2)-m(m+2)-n(n+4)\right)\end{array}\right.
\end{align}
Equating the eigenvalues (\ref{C2 short}) and (\ref{C2 susy}) will give a simple way of understanding the $k,f,H$-decomposition of
superconformal blocks.

Let us first discuss the short half-BPS blocks $\mathcal{B}_{0,\Delta}$. 
These are the only blocks for which the Casimir eigenvalue vanishes, and therefore the only blocks for which $k\neq 0$.
However, for each $R$-symmetry representation, both the chiral and reduced correlator can also contain a term
with vanishing Casimir eigenvalues: the functions $f_{0,m}(z,\a)$ and $H_{m,n,m,n}(z,\zb,a,\ab)$.
Looking at the disconnected OPE data $a_{m,n,m,n}^{(0)}$ and $b_{0,m}^{(0)}$ in eqs.~(\ref{Eq:a0pqqp}) and (\ref{Eq:a0chiral}),
we find that these come with unit coefficient for all the allowed $R$-symmetry representations.
On the other hand, only one half-BPS block, with double-trace dimension $\Delta=p+q$, is
expected physically to contribute to the disconnected correlator.
Consistency thus requires this half-BPS block to be equal to the sum over
all the vanishing-Casimir contributions to $f^{(0)}$ and $H^{(0)}$, and the non-identity contribution to $k$:
\be
\label{B multiplet}
\mathcal{B}_{0,\Delta}^{r,s} = \left\{\begin{array}{l}
\displaystyle
k=1, \qquad
f(z,\a)= \displaystyle \sum_{\substack{i=\max(|r|,|s|)}}^{\Delta-2}
k_{1+\frac{i}{2}}^{r,s}(z)\ k_{-\frac{i}{2}}^{-r,-s}(\a) ,
\\ 
\displaystyle
H(z,\bar{z},\alpha, \bar{\alpha})= \sum_{i=\max(|r|,|s|)}^{\Delta-4}\sum_{j=0}^{(\Delta-i)/2}
G_{j,i+j+4}^{r,s}(z,\bar{z})\ Z_{j,i+j}^{r,s}(\a,\ab).
\end{array}\right.
\ee
Since this holds for any $\Delta=p+q$, this must be equal to the half-BPS block for any $\Delta$.

For half-BPS blocks, a simple alternative is to use formulas from \cite{Dolan:2001tt} (see also appendix A of \cite{Basso:2017khq})
which express the blocks
as a finite sum of bosonic blocks, according to the finite list of primaries contained within the supermultiplet.
The formulas there are given for identical external operators but are easy to generalize,
since the list of primaries is the same and the coefficients can be fixed  from the Ward identity in eq.~\eqref{ward identity}.
Performing this exercise we obtain
\begin{align}
\mathcal{B}_{0,\Delta}^{r,s}&= G_{0,\Delta}^{r,s} Z_{0, \Delta}^{r,s} \nonumber \\
&+ \zeta(\Delta) G_{1,\Delta+1}^{r,s} Z_{1,\Delta-1}^{r,s} + \zeta(\Delta) \zeta(\Delta+2) G_{2,\Delta+2}^{r,s} Z_{0,\Delta-2}^{r,s} \nonumber \\
&+ \zeta(\Delta-2) \zeta(\Delta) G_{0,\Delta+2}^{r,s} Z_{2,\Delta-2}^{r,s} + \zeta(\Delta-2)\zeta(\Delta)\zeta(\Delta+2) G_{1,\Delta+3}^{r,s} Z_{1,\Delta-3}^{r,s} \nonumber \\
&+ \zeta(\Delta-2) \zeta^2(\Delta) \zeta(\Delta+2) G_{0,\Delta+4}^{r,s} Z_{0,\Delta-4}^{r,s},
\label{Eq:HalfBPS}
\end{align}
where $\zeta(\Delta) = \frac{(\Delta^2 - r^2)(\Delta^2-s^2)}{2^4 \Delta^2 (\Delta^2 -1)}$.
The block in the first line is always present and just has the quantum numbers of $\mathcal{B}_{0,\Delta}$.
The remaining lines account for superconformal descendants.
Because of the coefficient $\zeta(\Delta)$, the second line should only be kept for $\Delta \geq 2 + \min(|r|,|s|)$,
while the last two lines only contribute for $\Delta \geq 4+ \min(|r|,|s|)$. 
(For another approach to superconformal blocks, see \cite{Aprile:2017xsp}.)

Although eq.~\eqref{B multiplet} was derived using the disconnected correlator (where $r=s$), we find that this expression is precisely equal to eq.~(\ref{Eq:HalfBPS})
for the large set of values of $\Delta$ and $r\neq s$ that we have verified. We believe that this equality is an exact mathematical identity.

The analog of eq.~(\ref{B multiplet}) for semi-short and quarter-BPS blocks is discussed in appendix \ref{Appendix susy blocks}.
The fact that the data obtained from the inversion integral organizes
into a super-OPE with only double trace dimensions provides a very non-trivial consistency check.

\subsection{$\Delta^{(8)}$: A remarkable eighth-order differential operator} \label{Sec:useful operator}

The combination $\Delta^{(8)}$ in eq.~(\ref{Eq:Delta8}) will play an important role below.
It depends only on Casimir invariants like $h(h-1)$ and can in fact be viewed as a eight-order differential operator:
\be
 \Delta^{(8)} H_{\{p_i\}} \equiv
 \frac{z \zb\a\ab}{(z-\zb)(\a-\ab)} \left( \D_z -\D_\a\right)\left( \D_z -\D_\ab\right)\left( \D_{\zb} -\D_\a\right)\left( \D_{\zb} -\D_{\ab}\right)
 \frac{(z-\zb)(\a-\ab)}{z \zb\a\ab} H_{\{p_i\}}, \label{delta8 diff op}
\ee
where
\be \D_x \equiv x^2\partial_x(1-x)\partial_x -\tfrac12(r+s)x^2\partial_x -\tfrac14 rs x \label{Casimir}
\ee
with $r=p_{21}$, $s=p_{34}$.
Acting on the product $G_{\ell,\Delta+4}(z,\zb)Z_{m,n}(\a,\ab)$ this has precisely the eigenvalue in eq.~(\ref{Eq:Delta8}).
Its presence in the denominator of eq.~(\ref{Eq:a0pqqp}) suggests that $\Delta^{(8)}H^{(0)}$ is a simple function;
indeed by looking at individual cases we find that:
\be
 \Delta^{(8)} H^{(0)}_{pqqp} = \frac{u^{\frac{p+q}{2}+2}}{\sigma^{\frac{p+q}{2}-2}}
 \left(\frac{\tau^{q-2}}{v^{q+2}} + \delta_{p,q}\right)\times C(p)C(q) \label{Delta8 H0}.
\ee
This is precisely the form of a disconnected correlator of complex
scalar primaries with dimensions $p_i+2$ and $R$-symmetry $[0,p_i{-}2,0]$,
up to a normalization $C(p)=p^2(p^2-1)$, explaining the form of eq.~(\ref{Eq:a0pqqp}).

The operator $\Delta^{(8)}$ has two important properties.
First, it annihilates all the protected multiplets, as listed in table \ref{Tab:Multiplets}.
Second, $\Delta^{(8)} H_{\{p_i\}}$ generally transforms under $z\mapsto z/(z-1)$ crossing
like the correlator of scalar primaries with the mentioned quantum numbers,
which are precisely those of the superconformal descendant $\mathcal{L}^{p_i-2}\equiv Q^4\mathcal{O}^{p_i}$.
(The $\mathcal{L}^{p_i-2}$ are primaries with respect to bosonic isometries.)
We thus identify $\Delta^{(8)} H_{\{p_i\}}$ as the correlator of the product
$\mathcal{L}^{p_1-2}\mathcal{L}^{p_2-2}$ with its complex conjugate:
indeed this correlator has to annihilate protected multiplets since the first pair is proportional to $Q^8$.  
This generalizes the $p=2$ case studied in \cite{Drummond:2006by}, where $\mathcal{L}^{0}$ was the chiral Lagrangian density.

The correlators with different $p$'s will be combined into a single ten-dimensional dilaton amplitude in section \ref{sec:10D}.

\section{Tree-level correlator at strong coupling: order $1/c^{1}$} \label{Sec:Tree-level}

\begin{figure}
\centering
\fontsize{20pt}{23pt}\selectfont
\def\svgwidth{0.9\linewidth}
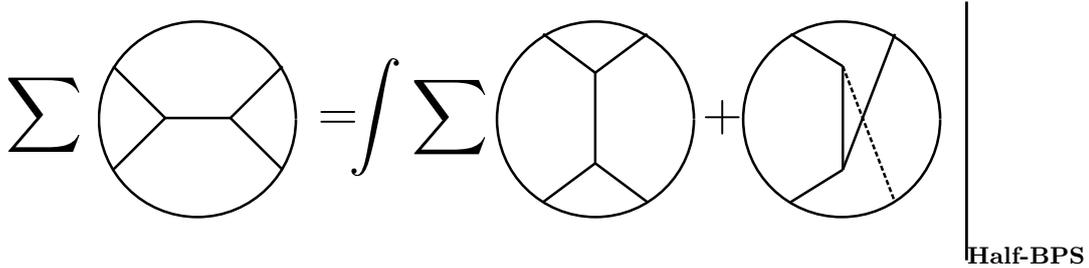
\caption{$s$-channel OPE data for single- and double-traces
can be reconstructed from cross-channel single-traces using the inversion formula.
In particular, demanding crossing symmetry unique fixes single-trace three-point coefficients.}
\label{Fig:crossing symmetry}
\end{figure}

At order $1/c$ in the limit of strong `t Hooft coupling, the correlator can be computed in principle in terms of tree-level
Witten diagrams in IIB supergravity on AdS${}_5\times$S${}_5$ \cite{Witten:1998qj}.
The present formalism offers a powerful alternative which relies only on CFT ideas.
The only input from supergravity is the assumption that all the single-trace operators are the half-BPS ones.
The double-discontinuity of the correlator comes exclusively from their exchange in the cross-channels,
which can be computed easily.
The inversion integral then reconstructs the full tree-level correlator, in the form of its OPE data,
as depicted in fig.~\ref{Fig:crossing symmetry}.

The double-discontinuity is computed explicitly by inserting in the cross-channel the half-BPS blocks
given in eq.~(\ref{Eq:HalfBPS}) or equivalently eq.~(\ref{B multiplet}), together with the crossing relation
\begin{equation}
G_{p_1p_2p_3p_4}(z,\bar{z},\alpha, \bar{\alpha}) = \left( \frac{z \bar{z}}{\alpha \bar{\alpha}} \right)^{\frac{p_1+p_2}{2}} \left( \frac{(1-\alpha)(1-\bar{\alpha})}{(1-z)(1-\bar{z})} \right)^{\frac{p_2+p_3}{2}} G_{p_3p_2p_1p_4}(1-z,1-\bar{z},1-\alpha,1-\bar{\alpha}).
\label{Eq:Crossing}
\end{equation}
If using eq.~(\ref{B multiplet}), it is important to note that
the different supersymmetry components $f,k,H$ in the decomposition (\ref{G ansatz}) mix under crossing: to find $H$ in one channel
one needs all of $f,k,H$ in the cross-channel.

\subsection{Fixing three-point coefficients from crossing symmetry}\label{ssec:3points}

For the correlator of eq.~(\ref{Eq:OriginalCorrelator}), the possible half-BPS states exchanged in the t-channel are
the representations of the form $\big[0,\Delta,0\big]$ in the
intersection $(\big[0,p_1,0\big] \times \big[0,p_4,0\big]) \cap (\big[0,p_2,0\big] \times \big[0,p_3,0\big])$.
From the tensor product formula in eq.~(\ref{Eq:TensorProductStates}) one finds the allowed range
\be
 {\rm max}(|p_1-p_4|,|p_2-p_3|) \leq \Delta \leq {\rm min} (p_1+p_4,p_2+p_3), \label{range half BPS}
\ee
where the allowed values of $\Delta$ differ from the endpoints by an even integer.

We will now demonstrate that, at order $1/c$, the three-point function of single-trace half-BPS operators is uniquely fixed by crossing symmetry,
to take the well-known value \cite{Lee:1998bxa}:
\be
 f_{pqr} = \frac{\sqrt{pqr}}{2\sqrt{c}}, \qquad |p-q|+2 \leq r \leq p+q-2. \label{f coeff}
\ee
The coefficients $f_{pqr}$ with $r=|p-q|$ or $r=p+q$ can be set to zero without loss of generality
by a suitable choice of basis for the ``single-trace'' operators. One simply chooses them to be
orthogonal to all multi-traces. This can always
be done by including suitable multiple of the latter in eq.~(\ref{def Op}), see for example \cite{Aprile:2018efk}.
(This assumes simply the existence of linearly independent  half-BPS double-trace operators
$[\mathcal{O}^p \mathcal{O}^q]$ for each $p,q\geq 2$, which follows from the $1/c^0$ analysis of the preceding section.
One also needs them to be linearly independent from the single traces $\mathcal{O}^{p+q}$,
which is physically expected at large $N$.)

Our proof of eq.~(\ref{f coeff}) uses only the chiral part of the correlator.
We illustrate the method starting with the simplest correlator, $f_{2222}$.
The only half-BPS block in the range (\ref{range half BPS}) which contributes to its double-discontinuity is the stress-tensor multiplet
$\Delta=2$, since the $\Delta=4$ block has double-trace dimensions and therefore no singularity.
This block is given as
\be
\mathcal{B}_{0,2}(z,\zb,\a,\ab) =
1+\frac{(z-\a) (z-\ab)\zb^2\log(1-z)-z^2 (\zb-\a) (\zb-\ab)\log(1-\zb)}{z\zb(\zb-z)\a \ab}.
\ee
Only this block and the identity contribute to the double discontinuity of $f$. 
We insert them into the crossing relation (\ref{Eq:Crossing}), extract the $f$ part by setting $\alpha=z$ (see eq.~(\ref{kf from G})),
and focus on the polar terms as $\zb\to 1$:
\be\begin{aligned}
\dDisc f_{2222}(\zb,\ab) &=  \dDisc\left[\frac{1-\ab}{(1-\zb)^2\ab} + \frac{3\ab-2}{(1-\zb)\ab}
+ f_{222}^2 \frac{1+\ab}{2(1-\zb)\ab}\right]
\\
&=  \dDisc\left[\frac{1}{y}-\frac{1}{2y^2} + \frac{3f_{222}^2}{4y}\right]k_0^{(0,0)}(\ab)
+ \dDisc\left[\frac{1}{y^2}+ \frac{f_{222}^2}{2y}\right]k_{-1}^{(0,0)}(\ab),
\end{aligned}
\ee
where on the second line $y=\frac{1-\zb}{\zb}$ and we have decomposed the $\ab$ dependence into eigenfunctions.
Plugging this into the inversion integral (\ref{f Inversion}) then gives the \emph{full} OPE decomposition for the chiral part:
\be \label{OPE f2222}
f_{2222}(\zb,\ab) = \sum_{j=0}^\infty \Big(1+(-1)^j\Big)
\left(\begin{array}{l}
 \displaystyle \left(\frac{(1{-}j)(j{+}2)}{2} + \frac{3f_{222}^2}{4}\right) r^{(0,0)}_{j+1} f_{j,0}(\zb,\ab)\\\displaystyle 
 + \left( (j{+}1)(j{+}2)+\frac{f_{222}^2}{2}\right)r^{(0,0)}_{j+2}f_{j,2}(\zb,\ab)\end{array} \right),
\ee
where $f_{j,m}$ is the function defined in eq.~(\ref{C2 susy}), and the $(-1)^j$ terms account for the $u$-channel contribution
(obtained generally by permuting operators $1$ and $2$, which does nothing here).
This OPE can be easily resummed, as discussed shortly.
However, what is most important here is to project it into its half-BPS contributions, which is simply the $j=0$ term:
\be\begin{aligned}
f_{2222}\Big|_{\rm 1/2-BPS} &= \left(2+ \frac{3f_{222}^2}{2}\right) f_{0,0}(\zb,\ab) +
\left(2+ \frac{f_{222}^2}{2}\right) f_{0,1}(\zb,\ab)
\\ &= f_{222}^2 \mathcal{B}_{0,2}  + \left(2+\frac{f_{222}^2}{2}\right) \mathcal{B}_{0,4},
 \label{f2222 crossing}
\end{aligned}\ee
where we have used from eq.~(\ref{B multiplet}) that the $f$ part of
$\mathcal{B}_{0,2} = f_{0,0}$ and $\mathcal{B}_{0,4} = f_{0,0}+f_{0,1}$.

The second line of eq.~(\ref{f2222 crossing}) is the main result of this exercise.
It shows that, if one inserts the stress tensor block $\mathcal{B}_{0,2}$ in the cross-channels with coefficient $f_{222}^2$,
the resulting double-discontinuity forces it to reappear in the $s$-channel with the same coefficient.
Thus crossing can be satisfied for any value of $f_{222}$.
For the present purposes, its value, $f_{222}=\sqrt{2/c}$ from eq.~(\ref{f coeff}), is simply a definition of the central charge $c$.
The double-trace block $\mathcal{B}_{0,4}$ then has a fixed coefficient.

We now proceed similarly for the $2323$ correlator, and find
that the double-discontinuity of the $t$- and $u$-channel OPE force its half-BPS part to be
\be
f_{2323}\Big|_{\rm 1/2-BPS} =
 \left( \frac13f_{222}f_{233} + \frac{7}{9} f_{233}^2\right)\mathcal{B}_{0,3}
+ \left(1+ \frac13 f_{222}f_{233} + \frac19 f_{233}^2\right)\mathcal{B}_{0,5}+O(1/c^2). \label{f2323 crossing}
\ee
Focusing on the first term, which is the only single-trace block,
the $s$-channel OPE is satisfied only if the coefficient of $\mathcal{B}_{0,3}$ is $f_{233}^2$.
Thus crossing symmetry requires that
\be
 f_{233}\left(\frac{1}{3}f_{222} - \frac{2}{9}f_{233}\right) =O(1/c^2), \label{f233 eq}
\ee
which has a nontrivial solution, $f_{233}=\frac{3}{2}f_{222}$, in precise agreement with eq.~(\ref{f coeff}).
The trivial solution $f_{233}=0$ could be ruled out using the stress tensor Ward identities:
the $\mathcal{O}^2$ multiplet contains the stress tensor, whose coupling to any pair of identical scalars
is nonzero and proportional to their dimension (consistent with the nontrivial solution).
More generally, the $2p2p$ crossing relation requires that
\be
 f_{2pp}\left(\frac{1}{p}f_{222} - \frac{2}{p^2}f_{2pp} \right) = O(1/c^2), \label{f2pp eq}
\ee
which has the nontrivial solution $f_{2pp} = p/2 f_{222}$ in agreement with both eq.~(\ref{f coeff}) and
the stress tensor Ward identity.
We find empirically that it seems possible to rule out the trivial solutions without relying on the Ward identities: for example,
setting $f_{233}=0$ (but not the other ones)
would eventually lead to a crossing equation with no solution for a higher correlator (namely $5566$).


We include $O(1/c^2)$ errors in the above, because double-trace operators,
such as semi-short ones, generally contribute to the double-discontinuity at this order. 

Further coefficients can be fixed by looking at more correlators. For example, the $3p3p$ correlators gives
two equations each, from the coefficients of the two single-trace blocks $\mathcal{B}_{0,p-1}$ and $\mathcal{B}_{0,p+1}$.
From these we can uniquely fix the $f_{3p\,p+1}$ and $f_{4pp}$ three-point functions for all $p$.
We find that the solution is always unique (up to an overall sign ambiguity for each operator: $\mathcal{O}^p\mapsto \pm \mathcal{O}^p$, which we fix by taking all three-point functions to be positive).

From the $4p4p$ correlators one similarly fixes all $f_{4p\,p+2}$ and $f_{6pp}$ three-point functions,
then from $5p5p$ with $p\geq 5$ one can fix $f_{5p\,p+1}$, $f_{5p\,p+3}$ and $f_{8pp}$, and so on.
There appears to be a simple pattern: the $qpqp$ family with a given $q$, once families with lower $q$'s have been used,
fixes all the $f_{qmn}$ coefficients in the range of eq.~(\ref{f coeff}) as well as $f_{2q{-}2\,pp}$.

We conclude that, under the assumption
that the only single-trace operators are the half-BPS ones (as expected at strong `t Hooft coupling),
crossing symmetry at order $1/c$ uniquely determines the three-point functions to have the values in eq.~(\ref{f coeff}) at this order.
A full analysis of the crossing relations at finite $c$ would be very interesting but is beyond our scope.
See \cite{Beem:2016wfs} for further results.

\subsection{Chiral (protected) correlator: deducing the free correlator} \label{ssec:free}

Interestingly, from this solution for the chiral correlator obtained at strong coupling (this is how we justified
having all single-traces operators being half-BPS),
we can recover Feynman diagrams of the free theory!

Consider again the protected correlator $f_{2222}$, given by the OPE in eq.~(\ref{OPE f2222}).
The infinite sum over $j$ can be computed easily if one puts in the expectation that the result is $1/(1-\zb)^2$ times a polynomial,
which can be found using a small number of terms. We obtain:
\be
 f_{2222}(z,\a) = \frac{z^2(1-\a)}{(1-z)^2\a} + \frac{z(2-z)}{1-z} + \frac{z^2}{\a}
 + \frac1c \frac{z(z+(3-2z)\a)}{\a(1-z)}\,. \label{f2222}
\ee
This does not look very suggestive but can be illuminated by comparing with the free theory.
Due to nonrenormalization theorems \cite{Seiberg:1993vc, Eden:2000bk, Bissi:2015qoa},
the protected quantity $f$ should be independent of the `t Hooft coupling and therefore
computable from the \emph{free theory} limit (small `t Hooft coupling) of the correlator.

In the free theory the correlator is a sum of Wick contractions which give products of $y_{ij}^2/x_{ij}^2$ factors.
In terms of cross-ratios (see eq.~(\ref{Eq:OriginalCorrelator})), the free correlator thus depends only on two variables:
$\mathcal{G}^{\rm free}_{pqrs}(u/\sigma, v/\tau)$.
Given that $f(z,\a)$ is itself a two-variable function, it is actually possible to go the other direction!
Indeed, we can solve eq.~(\ref{kf from G}) for the former in terms of $f$ and $k$:
\be
 \mathcal{G}^{\rm free}_{pqrs} =
 \frac{z-\a}{z\a}f_{pqrs}(z,\a)\Big|_{z= \frac{u}{\sigma} \frac{1-\frac{v}{\tau}}{\frac{u}{\sigma}-\frac{v}{\tau}},
\ \a=\frac{1-\frac{v}{\tau}}{\frac{u}{\sigma}-\frac{v}{\tau}}}
+  k_{pqrs} \left(\frac{u}{\sigma}\right)^{\frac12{\rm max}(|p_{12}|,|p_{34}|)}
  \left(\frac{\tau}{v}\right)^{\frac12{\rm max}(p_2+p_3-p_1-p_4,0)}. \label{Gfree from f}
\ee
The unit part $k$ can be computed generally by summing the
coefficients of all $\mathcal{B}_{0,\Delta}$ blocks in the OPE (including the double-trace one),
plus the identity block.  For example, eq.~(\ref{f2222 crossing}) gives $k_{2222}=3+\frac{3}{c}$.
Plugging the innocuous-looking eq.~(\ref{f2222}) into eq.~(\ref{Gfree from f}), we then get the free correlator:
\be
 \mathcal{G}_{2222}^{\rm free} = 1+\frac{u^2}{\sigma^2} + \frac{u^2\tau^2}{\sigma^2v^2}
 + \frac1c \left( \frac{u}{\sigma} + \frac{u\tau}{\sigma v} + \frac{u^2\tau}{\sigma^2v}\right).
\ee
This precisely matches with computing Wick contractions at weak coupling!
Note that this computation leveraged non-renormalization theorems
starting from the limit of \emph{strong} `t Hooft coupling.

It is straightforward to repeat these stops for other correlators, for example
\begin{align}
  \mathcal{G}^{\rm free}_{22pp} &= 1+ \frac1c\left(\frac{p}2 \frac{u}{\sigma} + \frac{p}2\frac{u\tau}{\sigma v} + p\frac{u^2\tau}{\sigma^2v}\right) +O(1/c^2)
  \qquad (p>2),
\\
 \mathcal{G}^{\rm free}_{3333} &= 1+\frac{u^2}{\sigma^2} + \frac{u^2\tau^2}{\sigma^2v^2}
+ \frac{9}{4c} \left(\frac{u^2}{\sigma^2}+ \frac{u\tau}{\sigma v}+ 2\frac{u^2\tau}{\sigma^2v}  + \frac{u^2\tau^2}{\sigma^2v^2}+ \frac{u^3\tau^2}{\sigma^3v^2} + \frac{u^3\tau}{\sigma^3v} \right) + O(1/c^2).
\end{align}
We found that these formulas agree precisely with the result of summing Wick contractions.
For $p\geq 4$ it is important to define external ``single-trace'' operators to be orthogonal to all double-trace ones,
reflecting the choice of basis stated below eq.~(\ref{f coeff}).
For the reader's convenience, a general formula for the free correlator at order $1/c$ is recorded in eq.~(\ref{Gfree Appendix}).

\subsection{Reduced correlator: Anomalous dimensions at order $1/c$}

Let us now turn to our main target, the correlator at order $1/c$ for external operators with general $R$-symmetry representations.
This gives information on
both the $1/c$ anomalous dimensions of double-trace operators and $1/c$ corrections to their OPE coefficients.

Our input again is the double-discontinuity written as a sum of half-BPS blocks with coefficients in eq.~(\ref{f coeff}),
and crossing relation (\ref{Eq:Crossing}). Explicitly, it is:
\be\begin{aligned}
\dDisc\,\mathcal{G}_{\{p_i\}}=
\dDisc\,\left( \frac{u}{\sigma}\right)^{\frac{p_1+p_2}{2}} \left( \frac{\tau}{v}\right)^{\frac{p_2+p_3}{2}}
\sum_{\Delta=\max(|p_{23}|,|p_{14}|)+2}^{\min(p_2+p_3,p_1+p_4)-2} \frac{\sqrt{p_1 p_2 p_3 p_4}\Delta}{4}
\mathcal{B}_\Delta^{p_{23},p_{14}}(z',\zb',\a',\ab') 
\label{Eq:CorrelatordDisc}
\end{aligned}\ee
where $x'=(1-x)$.
To get information on unprotected operators we first extract the $H$ part following the SUSY decomposition (\ref{G ansatz});
in the context of the $2222$ correlator (discussed at length in \cite{Alday:2017vkk}), this gives, for reference,
\be
\dDisc\, H_{2222}^{(1)} = \dDisc \left[\frac{u^3 (1-3u)}{(1-u)^3v} - \frac{2 u^5 \log u}{(1-u)^4v}\right]
\ee
where we have dropped all terms with no poles at $v=0$.\footnote{For unequal operators, one can generally
drop all powers of $v$ higher than $\frac12 {\rm min}(0,p_1+p_4-p_2-p_3)$, as these correspond to double-trace dimensions.
These vanish upon integration as can be seen from the zeros of the $1/\Gamma$ factors in eq.~(\ref{Eq:zbarIntegral}).}
We then plug into the Lorentzian inversion integral (\ref{Eq:cInversion 2}). 

Loosely speaking, the $\log u$ term informs us about anomalous dimensions, whereas the
constant term corrects in addition the OPE coefficients.  We begin with the former.
Converting to $x=\frac{z}{1-z}$ and $y=\frac{1-\zb}{\zb}$ variables, the above gives
\be
\dDisc\, \frac{\zb-z}{z \zb}H_{2222}^{(1)}\Big|_{\log u} = -2x^3\ \dDisc \frac{1}{y}
\ee
and the inversion integral (\ref{Eq:cInversion 2})
gives straightforwardly, using formulas (\ref{Eq:zbarIntegral}) and (\ref{Eq:zIntegral}) for the $x$ and $y$ powers,
\be
 \langle a^{(0)}\gamma^{(1)}(h,\hb,0,0) \rangle_{2222} = -2(h+1)(h)(h-1)(h-2) r^{0,0}_h r^{0,0}_\hb Z_{0,0}^{0,0}.
\ee
This is in agreement with \cite{Alday:2017vkk}. The bracket notation indicates that the result is generally the sum of multiple nearly-degenerate operators.
The set of correlators really gives
a matrix of anomalous dimensions matrix, whose eigenvalues are the actual anomalous dimensions.

The operators which can mix with each other have the same R-symmetry representation $[m,n-m,m]$
and twist $\tau=\Delta-\ell$.   To obtain their matrix element in the basis of disconnected theory operators,
we need to divide by the disconnected OPE data in eq.~(\ref{Eq:a0pqqp}),
and since $ \langle a^{(0)} \rangle_{pqqp} = \langle a^{(0)} \rangle_{qppq}$, we have:
\begin{equation}
\gamma^{(1)}_{pq,rs}\equiv \frac{\langle a^{(0)} \gamma^{(1)} \rangle_{pqrs} }{\sqrt{\langle a^{(0)} \rangle_{pqqp} \langle a^{(0)} \rangle_{rssr}}} = \frac{\langle a^{(0)} \gamma^{(1)} \rangle_{pqrs}^t + (-1)^{\ell+m} \langle a^{(0)} \gamma^{(1)} \rangle_{qprs}^t }{\sqrt{\langle a^{(0)} \rangle_{pqqp} \langle a^{(0)} \rangle_{rssr} }},
\label{Eq:anomDim}
\end{equation}
where the $t$ and $u$ superscripts denote the t and u-channel exchanges respectively. For a given twist and R-symmetry representation, one can find the list of pairs $(p,q)$, with $p\leq q$, for which double-trace
operators $[pq]_{\ell,\Delta,m,n}$ exist, and construct the matrix in eq.~(\ref{Eq:anomDim}).
Its eigenvalues are then the \textit{actual} OPE coefficients of the double-trace operators.

To perform the computation, we wish to organize our data in terms of twist $\tau$ and R-symetry representations labelled by $\big[m, n-m, m\big]$. One must also consider separately the cases where the spin $\ell$ is either even or odd. For a given twist and representation, one must find the possible scaling dimensions of the operators in the four-point function. One can simply consider the constraints on the three-point vertices. Given the possible $(pqrs)$ scaling dimensions and values of twist, R-symmetry Dynkin labels and spin, one can diagonalize the corresponding matrix.

We will change our notation to write the anomalous dimension matrix $\left( \gamma^{(1)} \right)_{\tau,m,n}^{\ell}$ labelled by the twist $\tau$, R-symmetry labels $m,n$ and spin $\ell= \{+,-\}$ corresponding to either even or odd spin. For example, let us consider the nontrivial representation $\big[0,2,0\big]$. The minimal twist associated with this representation is $\tau=6$. For odd spin, the only three-point vertex is given by external operators $(p,q)=(2,4)$, in which case it is straightforward to diagonalize. For even spin, one has $(2,4)$ and $(3,3)$. The corresponding matrix is 
\begin{equation}
\begin{pmatrix}
\dfrac{\langle a^{(0)} \gamma^{(1)} \rangle_{2442}^t + (-1)^{\ell} \langle a^{(0)} \gamma^{(1)} \rangle_{4242}^t }{\sqrt{\langle a^{(0)} \rangle_{2442} \langle a^{(0)} \rangle_{2442}} } & \dfrac{\langle a^{(0)} \gamma^{(1)} \rangle_{2433}^t + (-1)^{\ell} \langle a^{(0)} \gamma^{(1)} \rangle_{4233}^t}{\sqrt{\langle a^{(0)} \rangle_{2442} \langle a^{(0)} \rangle_{3333}} } \\
\dfrac{\langle a^{(0)} \gamma^{(1)} \rangle_{3342}^t+ (-1)^{\ell} \langle a^{(0)} \gamma^{(1)} \rangle_{3342}^t}{\sqrt{\langle a^{(0)} \rangle_{3333} \langle a^{(0)} \rangle_{2442}} } & \dfrac{\langle a^{(0)} \gamma^{(1)} \rangle_{3333}^t+ (-1)^{\ell} \langle a^{(0)} \gamma^{(1)} \rangle_{3333}^t}{\sqrt{\langle a^{(0)} \rangle_{3333} \langle a^{(0)} \rangle_{3333}} }
\end{pmatrix}.
\end{equation}

Consider the $2442$ component. One finds that the only possible R-symmetry representation is $\big[0,2,0\big]$. The reduced correlator inversion integral yields
\begin{align}
c(\ell,\Delta,0,2) &=
\int_0^1 \frac{dz}{z^2}(1-z)^2 k_{1-h}^{2,2}(z) z^{-1} \int_0^1 \frac{d\bar{z}}{\zb^2} (1-\zb)^2 \kappa(\hb) k_{\hb}^{2,2}(\zb) \zb^{-1} \nonumber \\
& \times \left[2 \left( \frac{z}{1-z} \right)^3 - 4 \left( \frac{z}{1-z} \right)^4 - 4 \left( \frac{z}{1-z} \right)^5 \log(z) \right]
\dDisc\left( \frac{\zb}{1-\zb} \right)^3 
\end{align}
where we have omitted non-singular terms.
This formula contains corrections to the OPE coefficients $a^{(1)}$ and anomalous dimensions $\gamma^{(1)}$.
Focusing now on the latter, which are related to double-poles with respect to $h$, they can be obtained
simply by integrating the coefficient of $\log(z)$, see eq.~\eqref{x log(z) integral}.

Using eqs.~\eqref{Eq:zbarIntegral}-\eqref{Eq:zIntegral}, the above matrix with $h=1+\frac{\tau}{2}$ gives
\be
 \left(\gamma^{(1)}\right)^+_{6,0,2}=\frac{-60}{(\hb-4)(\hb-1)(\hb)(\hb+3)}
 \begin{pmatrix}
 12+\hb(\hb-1) & 6\sqrt{\hb(\hb-1)} \\
  6\sqrt{\hb(\hb-1)} & 6+\hb(\hb-1)
 \end{pmatrix}   \label{02 matrix}
\ee
which has the two eigenvalues:
\begin{equation}
\lambda_{6,0,2}^+ = \left\{-\frac{60(\hb+2)}{(\hb-4)(\hb-1)(\hb)},\
-\frac{60(\hb-3)}{(\hb-1)(\hb)(\hb+3)}\right\}
= \Big\{\frac{-\Delta_{0,2}^{(8)}(4,\hb)}{(\hb-4)_6},\ \frac{-\Delta_{0,2}^{(8)}(4,\hb)}{(\hb-2)_6}\Big\}
\end{equation}
where $(...)_6$ is the Pochhammer symbol and $\Delta^{(8)}$ is the polynomial introduced in eq.~(\ref{Eq:Delta8}).
This way of writing the result was suggested by a recent conjecture by \cite{Aprile:2018efk}, discussed shortly.
For $\tau=8$, again for the $\big[0,2,0\big]$ representation, the matrix is somewhat bigger:
\begin{equation}
\begin{pmatrix}
(2442) & (2433) & (2435) & (2444) \\
(3342) & (3333) & (3335) & (3344) \\
(3542) & (3533) & (3553) & (3544) \\
(4442) & (4433) & (4453) & (4444)
\end{pmatrix}.
\end{equation}
For odd spin, one removes the second and last rows and columns.
We then find the eigenvalues, for even and odd spins:
\be\begin{aligned}
 \lambda_{8,0,2}^- &= -\Delta_{0,2}^{(8)}(5,\hb) \times \Big\{ \frac{1}{(\hb-4)_6},\ \frac{1}{(\hb-2)_6} \Big\},
\\
 \lambda_{8,0,2}^+ &= -\Delta_{0,2}^{(8)}(5,\hb) \times \Big\{ \frac{1}{(\hb-5)_6},\ \frac{1}{(\hb-3)_6},\ \frac{1}{(\hb-3)_6},\ \frac{1}{(\hb-1)_6} \Big\}.
\end{aligned}\ee
Note that in the even spin case, the second and third eigenvalues are degenerate.

It is amazing that the eigenvalues of nontrivial matrices
give rational functions of $\hb$.

Reciprocity symmetry, $\hb\to 1-\hb$, is preserved in an interesting way.
The matrix elements in eq.~(\ref{02 matrix}) are all invariants under $\hb\to 1-\hb$
(this is always true and trivially follows from the form of the integrals (\ref{Eq:zbarIntegral})-(\ref{Eq:zIntegral})).
The individual eigenvalues are not invariant, but the \emph{set} of eigenvalues is, as it should:
reciprocity interchanges the first and last eigenvalue.

One can continue this process to obtain anomalous dimension eigenvalues in different R-symmetry
representations and higher twist which would involve diagonalizing larger dimensional matrices.
A conjecture for the outcome was formulated recently in \cite{Aprile:2018efk} (who obtained it by OPE-decomposing
a Mellin-space formula for the general $pqrs$ correlators previously conjectured in \cite{Rastelli:2016nze}).
These authors conjectured that the eigenvalues always take the form:
\begin{equation}
\gamma^{(1)}_{pqrs} (\tau,\hb,\ell, m,n) = - \frac{1}{c} \frac{\Delta_{m,n}^{(8)}(1+\frac{\tau}{2},\hb)}{\left(\ell+m+1 + \nu \right)_6}
\label{Eq:Anomalous Dimension}
\end{equation}
where $\ell= \hb - \frac{\tau}{2}-2$ is the angular momentum and $\nu$ is an integer shift
that distinguishes different eigenvalues of the matrix.

Generically, diagonalizing the anomalous dimension matrix will yield degenerate states. This degeneracy can be worked out experimentally.
We find that the allowed shifts are
\begin{align}
\nu &= \displaystyle \frac{1-(-1)^{\ell+m}}{2}+2u, \quad 0\leq u\leq \tfrac{1}{2}(\tau-2m-5+ \delta) \label{multiplicity shift} \\
\delta &= \begin{cases}
(-1)^{\ell+m} & \text{if} \mod(n-m,2)=0 \\
0 & \text{otherwise}.
\end{cases}
\end{align}
The multiplicity for a given shift is then
\begin{equation}
\text{min}\left(u+1, \tfrac{1}{2} \left(n-m+1+ \delta \right), \tfrac{1}{2}\left( \tau-2m-3+\delta \right)-u \right).
\label{multiplicity}
\end{equation}
The first case sets the degeneracy for the 
first few cases, which  increases incrementally from 1 until it reaches its maximum multiplicity given by the second condition.
If the twist is large enough, the multiplicity will plateau until the third condition becomes relevant, after which it decreases symmetrically.
In all cases, the multiplicities add up to the dimension of the mixing matrix
(which is equal to the number in the second condition, times $\frac{\tau-m-n-2}{2}$).
 
\begin{figure}[h]
\centering
\begin{subfigure}[b]{0.45\textwidth}
\includegraphics[width=\textwidth]{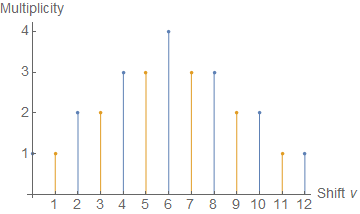}
\end{subfigure}
\begin{subfigure}[b]{0.45\textwidth}
\includegraphics[width=\textwidth]{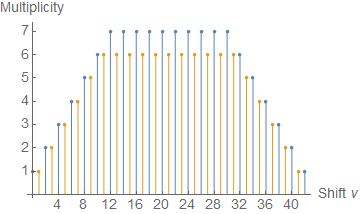}
\end{subfigure}
\caption{Even and odd spins are in blue and orange respectively. On the left is the multiplicity for $\tau=20$ and on the right is the multiplicity for $\tau=50$ of the same R-symmetry representation. Given eq.~\eqref{multiplicity}, the maximal multiplicity for [2, 12, 2] is 7 and 6 for even and odd spin respectively.}
\label{Fig:multiplicity}
\end{figure}

Consider for example the $\big[2,12,2\big]$ representation with twist up to $\tau=20$ and $\tau=50$. One can find the multiplicity of the degenerate states in fig.~\ref{Fig:multiplicity}.
This matches with the finding in ref.~\cite{Aprile:2018efk}, where the allowed states were nicely organized into a $45^\circ$-rotated rectangle
in the $(p,q)$ plane.

From this analysis, we see that the Pochhammer term serves as a useful quantity to resolve the double trace mixing problem. This observation will be recycled in the next section.

\section{An accidental 10-dimensional conformal symmetry} \label{sec:10D}

In this section we propose an explanation for the remarkable conjecture (\ref{Eq:Anomalous Dimension}):
that it originates from the conformal flatness of the  AdS${}_5\times $S${}_5$ geometry
and an accidental SO(10,2) conformal symmetry of the supergravity four-point amplitude.
This will lead to new conjectures which we will test.

\subsection{A mysterious equality with 10-dimensional $S$-matrix partial waves}

Our main inspiration will be the following empirical observation: the conjectured
anomalous dimension is equal to a quantity computed in \emph{flat} ten-dimensional space.
We consider the $2\to 2$ scattering of identical complex axi-dilatons in IIB supergravity:
\be
A_{10}^{\rm tree} = \frac{8\pi G_N\delta^{16}(Q)}{stu} \longrightarrow
8\pi G_N\frac{s^4}{stu} .
\label{Eq:10D-amplitude}
\ee
Below we'll use that $8\pi G_N=\frac{\pi^5L^8}{c}$ with $c=\frac{N^2-1}{4}$ 
in the AdS${}_5\times $S${}_5$ context, to express this in terms of $c$ and the AdS radius $L$.
We now consider the partial-wave decomposition of this amplitude;
in general dimension this can be written using Gegenbauer polynomials (see for example \cite{Giddings:2007qq}) as:
\begin{equation} \label{S matrix partial waves}
i A_{10}(s, \cos\theta) =
\frac{2^{d} \pi^{\frac{d-2}{2}}}{s^{\frac{d-4}{2}}\Gamma\big(\frac{d-2}{2}\big)}
\displaystyle \sum_{\ell>0, \text{ even}} (\ell+1)_{d-4}(2\ell+d-3) A_\ell(s) \ \tilde{C}^{(\frac{d-3}{2})}_\ell(\cos\theta) ,
\end{equation}
where
$\tilde{C}^{(\lambda)}_\ell(x)= C^{(\lambda)}_\ell(x)/C^{(\lambda)}_\ell(1)$ is a Gegenbauer polynomial normalized to unity at 1,
the scattering angle is $\cos(\theta)=1+\frac{2t}{s}$, and we should put $d=10$ in the above.
The normalization factors are such that $A_\ell(s)=1$ for the disconnected part of the $S$-matrix of two identical particles.
Retaining the disconnected part and computing the $A_\ell$ using the orthogonality relation\footnote{
Specifically, the relevant relation is:
\be A_\ell(s) = 1 + \frac{i s^3}{786432 \pi^4} \int_0^\pi d\theta  \sin^7(\theta) \ \tilde{C}^{(7/2)}_\ell(\cos\theta) \ A_{10}(s, \cos\theta).\ee
}
we find:
\be
A_\ell(s) = 1 + i \frac{\pi}{c} \frac{\left(L\sqrt{s}/2\right)^8}{(\ell+1)_6}\,. \label{Eq:10D-partial-wave}
\ee
This should be compared with the exponential of the anomalous dimension in eq.~(\ref{Eq:Anomalous Dimension}),
which controls the phase of each conformal block in the bulk point limit (see for example \cite{Heemskerk:2009pn,Maldacena:2015iua}):
\be
 e^{-i\pi\gamma} = 1 + i\frac{\pi}{c} \frac{\Delta^{(8)}}{(\ell_{\rm eff}+1)_6} + \mathcal{O}(1/c)\,. \label{Eq:4D-partial-wave}
\ee
The resemblance of the phases (\ref{Eq:10D-partial-wave}) and (\ref{Eq:4D-partial-wave}) is too striking to be an accident
and this suggests a direct relation between the four-dimensional theory and the \emph{flat} ten dimensional supergravity.
We will not fully derive such a relation from first principles here, but we will try to guess what form it could take
and deduce precise implications.

\subsection{Statement of the conjecture}

To explain how an observable of 10-dimensional supergravity on AdS${}_5\times$S${}_5$
might be entirely determined by the $S$-matrix of the same theory on \emph{flat} space, we propose
a scenario based on the following two observations:
\begin{itemize}
\item The AdS${}_5\times$S${}_5$ metric is conformally equivalent to flat space.
To check this, write the 10-dimensional flat space metric
using radial and angular variables for the 6 extra dimensions,
with $d\Omega_5^2$ is the metric on $S_5$:
\be\begin{aligned}
 ds^2_{\rm Minkowski} &= \sum_{\mu=0}^3 dx^\mu dx_\mu + dr^2 + r^2 d\Omega_5^2
 \\ &= r^2 \times \left( \frac{dx^\mu dx_\mu + dr^2}{r^2} + d\Omega_5^2\right)
  \equiv \frac{r^2}{L^2} \times ds^2_{{\rm AdS}{}_5\times {\rm S}{}_5}.
\end{aligned}\ee
\item By a happy coincidence, the combination $G_N \delta^{16}(Q)$ is dimensionless.
Divided by this ``coupling'', the four-point tree amplitude (\ref{Eq:10D-amplitude})
is in fact conformally invariant:
\be
K_\mu \frac{\delta^{10}(\sum_i p_i)}{(p_1{\cdot}p_2)(p_1{\cdot}p_3)(p_2{\cdot}p_3)} =0.
\ee
This can be checked using the conformal generators in momentum space,
\be K_\mu = \sum_{i=1}^4 \left(
\frac{p_{i\mu}}2  \frac{\partial}{\partial p_i^\nu}\frac{\partial}{\partial p_{i\nu}}
-p_i^\nu \frac{\partial}{\partial p_i^\nu}\frac{\partial}{\partial p_{i}^\mu}
-\frac{d-2}{2} \frac{\partial}{\partial p_i^\mu}\right).
\ee
\end{itemize}
In effect, we will now play a game where we pretend that IIB supergravity is a conformal theory
and work out the implications for four-point correlators.
In particular, we should find an action of the ten-dimensional conformal group SO(10,2).

This SO(10,2) contains as a subgroup the SO(4,2)$\times$ SO(6) bosonic isometries of AdS${}_5\times$S${}_5$,
so a natural action can be defined by building 12-vectors $w_i$:
\be
w_i\equiv (X_i,y_i).
\ee
Here $X_i$ is the embedding coordinate corresponding to the space-time point $x_i^\mu$, on which SO(4,2) acts,
and $y^i$ is the null 6-vector introduced in eq.~(\ref{cross-ratios sigma tau}) and on which the SO(6) $R$-symmetry acts.

Motivated by the above, we conjecture that all tree-level four-point correlators arise from a single
SO(10,2)-invariant object.

To state this more explicitly we consider
the correlator of Lagrangian insertions $\mathcal{L}^{p_i-2}$ defined in section \ref{Sec:useful operator} (which are susy descendent
$Q^4 \mathcal{O}^{p_i}$ of scaling dimension $2+p_i$), since these can be easily identified with axi-dilatons in ten dimensions.
Indeed, the simplest case, $p_i=0$, has scaling dimension 4 which is
the correct value for a free field in ten dimensions.
The operators with $p_i> 0$ are then identified to be its S${}_5$ spherical harmonics.
To extract a given spherical harmonic we simply need to extract the component of a ten-dimensional
expression with the correct homogeneity degree in the null vectors $y_i$:
\def\tp#1{{\tilde{p}_{#1}}}
\be
\langle0| \mathcal{L}^{\tp{1}}(x_1) \mathcal{L}^{\tp{2}}(x_2) \mathcal{L}^{\tp{3}}(x_3) \mathcal{L}^{\tp{4}}(x_4)|0 \rangle =
\langle \phi(w_1) \phi(w_2) \bar{\phi}(w_3) \bar{\phi}(w_4) \rangle_{10}
\Big \rvert_{y_1^{\tp{1}}y_2^{\tp{2}}y_3^{\tp{3}}y_4^{\tp{4}}}\,, \label{LLLL from 10D}
\ee
with $\tp{i}=p_i-2$.
The expectation value on the right is formal since
we haven't defined a 10D CFT  (and won't define one).
It simply denotes a function of the four points $w_i$ which transforms like a 10D CFT correlator
of four scalars with scaling dimension 4, that is:
\be
\langle \phi(w_1) \phi(w_2) \bar{\phi}(w_3) \bar{\phi}(w_4) \rangle_{10} \equiv
\frac{G_{10}(u_{10},v_{10})}{\big((x_{12}^2-y_{12}^2)(x_{34}^2-y_{34}^2)\big)^4} \label{10d four scalars}
\ee
where
\be
u_{10} \equiv \frac{(x_{12}^2-y_{12}^2)(x_{34}^2-y_{34}^2)}{(x_{13}^2-y_{13}^2)(x_{24}^2-y_{24}^2)},\qquad
v_{10} \equiv \frac{(x_{23}^2-y_{23}^2)(x_{14}^2-y_{14}^2)}{(x_{13}^2-y_{13}^2)(x_{24}^2-y_{24}^2)}
\ee
are the then-dimensional cross-ratios.

\begin{figure}[h]
\centering
\def\svgwidth{\linewidth}
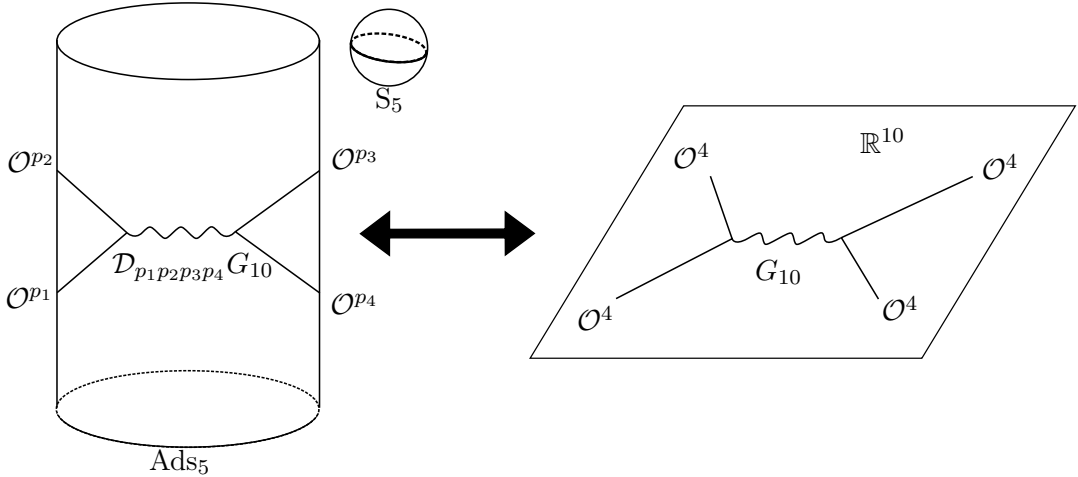
\caption{The flat-space limit of a scattering process in AdS${}_5 \times$ S${}_5$ map to one in $\mathbb{R}^{10}$. In the flat-space limit, the dilaton has $\Delta=4$. We conjecture that the ten-dimensional dilaton correlator $G_{10}$ can be projected to a AdS${}_5 \times$ S${}_5$ correlator by a suitable differential operator $\mathcal{D}_{p_1 p_2 p_3 p_4}$ acting on $G_{10}$.}
\label{Fig:AdS flat space}
\end{figure}

Let us now write the prediction in eq.~(\ref{LLLL from 10D}) in terms of cross-ratios,
by dividing the left-hand-side by the factor in eq.~(\ref{Eq:OriginalCorrelator}) (with the $p_i$ shifted by $\pm 2$ for the
$x$ and $y$ dependence, respectively).
We can extract the component of correct homogeneity degree in $y$ by
performing contour integrals in auxiliary variables $a_i$
introduced via the rescaling $y_{ij}^2\equiv y_i\cdot y_j \mapsto a_i a_j y_{ij}^2$.
With a suitable rescaling of the $a$'s this gives a formula with cross-ratios only:
\begin{align}
\tilde{H}_{p_1 p_2 p_3 p_4}(u,v,\sigma,\tau) &= 
\oint \prod_{i=1}^4 \left[\frac{da_i\,a_i^{1-p_i}}{2\pi i} \right]\ \frac{(u/\sigma)^{\frac{p_1+p_2}{2}-2}}{(1- \frac{\sigma}{u} a_1 a_2 )^4 (1- a_3 a_4)^4}
\nonumber\\&\quad \times G_{10}
\left(u\frac{(1-\frac{\sigma}{u} a_1 a_2)(1- a_3 a_4)}{(1- a_1 a_3)(1-a_2 a_4)}, v\frac{(1-\frac{\tau}{v} a_2 a_3)(1 - a_1 a_4)}{(1 -a_1 a_3)(1- a_2 a_4)} \right)
\nonumber\\ &\equiv \mathcal{D}_{p_1p_2p_3p_4}G_{10}(u,v).\label{Generating Function 10-4D}
\end{align}
The last line emphasizes the fact that the contour integral really just defines a differential operator.
In the first few cases for example we find:
\be\begin{aligned}
\mathcal{D}_{2222} &= 1, \label{D examples} \\
\mathcal{D}_{2332} &= -\frac{\sqrt{u}}{\sqrt{\sigma}} \tau \partial_v,\\
\mathcal{D}_{2233} &= 4 - u \partial_u, \\
\mathcal{D}_{3333} &= 16  -8 u\partial_u + \frac{u+\sigma}{\sigma} (u\partial_u)^2 +2\frac{u}{\sigma} u\partial_u v\partial_v
+ \frac{u(v+\tau)}{\sigma v} (v\partial_v)^2\,.
\end{aligned}\ee
Relation (\ref{Generating Function 10-4D}) is the main result of this section: it expresses
all tree-level correlators as derivatives of a single one, $G_{10}=\tilde{H}_{2222}$, the stress-tensor multiplet correlator.
This is depicted in fig.~\ref{Fig:AdS flat space}.

It remains to precisely define the left-hand-side $\tilde{H}_{2222}$.  The correct guess
is of course to be ultimately justified by explicit computations,
as done in the next subsection, but it may be helpful to handwave-motivate the result here.
Since the spherical harmonics of the ten-dimensional dilaton most naturally map to correlators of $\mathcal{L}^{p_i}$,
the most natural guess would be $\tilde{H}=\Delta^{(8)}H$ as defined in section \ref{Sec:useful operator}.
\emph{However}, to reveal its hidden conformal symmetry, we had to divide the 10D tree amplitude by the dimensionless
``coupling'' $G_N\delta^{16}(Q)/16$, which eqs.~(\ref{Eq:10D-partial-wave}) and (\ref{Eq:4D-partial-wave}) suggest to identify
as a multiple of  $\Delta^{(8)}/c$.  Thus a better guess is to divide the order $1/c$ dilaton
correlator by $\Delta^{(8)}$ and thus consider $H$ itself, which turns out to be a correct prediction.
We have to bear in mind that this division is ill-defined however unless one cancels the protected contributions killed by $\Delta^{(8)}$;
a simple work-around is to subtract the free correlator.
The correct guess for the functions $\tilde{H}$ which satisfies eq.~(\ref{Generating Function 10-4D}), including a normalization factor,
thus turns out to be:
\be
 \tilde{H}^{(0)}_{pqrs} \equiv \frac{4}{\sqrt{p_1p_2p_3p_4}}\Delta^{(8)} H^{(0)}_{pqrs}, \qquad
 \tilde{H}^{(1)}_{pqrs} \equiv \frac{4}{\sqrt{p_1p_2p_3p_4}}\left(H^{(1)}_{pqrs}-H^{(1)\rm free}_{pqrs}\right).  \label{def Htilde}
\ee
If this sequence were to continue at loop level supergravity,
a natural form of the next term would be
$\tilde{H}^{(2)}_{pqrs}=\frac{1}{\Delta^{(8)}}(H^{(2)}_{pqrs}-H^{(2)\rm free}_{pqrs})$, but we leave this for further study.

Eqs.~(\ref{Generating Function 10-4D}) and (\ref{def Htilde}), independent of their origins,
yield precise and testable relations among tree-level correlators.

\subsection{Checking the conjecture from the calculated correlators} \label{ssec:tests}

Let us first check the relation for the disconnected correlator $\tilde{H}^{(0)}$.
According to eq.~(\ref{Delta8 H0})
\be
 G_{10}^{(0)}(u,v)= \Delta^{(8)} H^{(0)}_{2222} = C(2)^2 \left( u^4+\frac{u^4}{v^4}\right), \label{Htilde 2222 free}
\ee
with $C(p)=p^2(p^2-1)$.
The formula in eq.~(\ref{Generating Function 10-4D}) then predict the other correlators as
\be\begin{aligned}
\tilde{H}^{(0)}_{p_1p_2p_3p_4} &= 
\frac{u^{\frac{p_1+p_2}{2}+2}}{u^{\frac{p_1+p_2}{2}-2}}
\oint \prod_{i=1}^4 \left[\frac{da_i\,a_i^{1-p_i}}{2\pi i} \right]
 \left(
 \frac{1}{(1- a_1 a_3)^4(1-a_2 a_4)^4} + \frac{1}{(v-\tau a_2 a_3)^4(1 - a_1 a_4)^4}
 \right)
\\ &=
\frac{4}{p_1p_2}C(p_1)C(p_2) \frac{u^{\frac{p_1+p_2}{2}+2}}{\sigma^{\frac{p_1+p_2}{2}-2}}
 \left(
 \delta_{p_1,p_3}\delta_{p_2,p_4}+ \delta_{p_1,p_4}\delta_{p_2,p_3} \frac{\tau^{p_2-2}}{v^{p_2+2}}\right),
\end{aligned}\ee
where we have used the integral $\oint \frac{6 da\ a^{1-p}}{2\pi i(1-a)^4} = C(p)/p$.
This is in precise agreement with eq.~(\ref{Delta8 H0}), a first sanity check.
Essentially this just fixes the normalization factors $\sqrt{p_i}$ in eq.~(\ref{def Htilde}).

The real test now is the tree-level correlator $\tilde{H}^{(1)}$.
The formula in eq.~(\ref{Generating Function 10-4D}) predicts the higher correlators as specific derivatives
of the first instance \cite{Arutyunov:2000py, Dolan:2006ec},\footnote{
Explicitly, $\bar{D}_{2,4,2,2}(u,v)=\partial_u\partial_v(1+u\partial_u +v\partial_v)
\frac{2{\rm Li}_2(z)-2{\rm Li}_2(\zb)+ \log(z\zb)\log\frac{1-z}{1-\zb}}{z-\zb}$. }
\be
 G_{10}^{(1)}(u,v) = \tilde{H}^{(1)}_{2222} = -u^4\bar{D}_{2,4,2,2}(u,v).
 \label{H2222 Dbar}
\ee
Acting with $\mathcal{D}_{p_1 p_2 p_3 p_4}$ on this gives predictions for the correlators of spherical harmonics.
It is instructive to compare with the literature.
For some of the cases presented in eqs.~\eqref{D examples}, refs.~\cite{Dolan:2004iy, Dolan:2006ec,Berdichevsky:2007xd,Uruchurtu:2008kp,Aprile:2017xsp} provided explicit results in terms of $\bar{D}$ functions. Using derivative relations to generate higher weight $\bar{D}$ functions (see e.g. ref.~\cite{Arutyunov:2002fh}), we evaluated these results as functions of $z,\bar{z}$ variables and compared with $\tilde{H}^{(1)}_{p_1 p_2 p_3 p_4}$ as computed from the action of the differential operator $\mathcal{D}_{p_1 p_2 p_3 p_4}$ on eq.~\eqref{H2222 Dbar}.
We find agreement with the literature up to a normalization constant:
\be\begin{aligned}
\tilde{H}_{2233}^{(1)} &= -6 u^5 \bar{D}_{3,5,2,2}, \label{H1 Comparison} \\
\tilde{H}_{3333}^{(1)} &= -9 u^5 \left( (1+\sigma + \tau) \bar{D}_{3,5,3,3} + \left( \bar{D}_{2,5,2,3} + \sigma \bar{D}_{3,5,2,2} + \tau \bar{D}_{2,5,3,2} \right) \right) , \\
\tilde{H}_{4444}^{(1)} &=-16 u^6 \Big( \left(\sigma+ \tau+ \sigma \tau + \frac{1}{4}(1+\sigma^2 + \tau^2) \right) \bar{D}_{4,6,4,4}  \\
& \ - \left( \sigma \bar{D}_{4,6,2,4} + \tau \bar{D}_{2,6,4,4} + \sigma \tau \bar{D}_{4,6,4,2} \right) \\
& \ + 2 \left(\sigma \bar{D}_{3,6,2,3} + \tau \bar{D}_{2,6,3,3} + \sigma \tau \bar{D}_{3,6,3,2} \right) \\
& \ + \frac{1}{2} \left(\bar{D}_{2,6,2,4} + \sigma^2 \bar{D}_{4,6,2,2} + \tau^2 \bar{D}_{2,6,4,2} \right) \\
& \ + \frac{1}{2} \left(\bar{D}_{3,6,3,4} + \sigma^2 \bar{D}_{4,6,3,3} + \tau^2 \bar{D}_{3,6,4,3} \right) \Big) \,.
\end{aligned}\ee
We stress that in those papers, the result came from explicit supergravity or Mellin-space calculations whereas our result came from the action of the differential operator $\mathcal{D}_{p_1 p_2 p_3 p_4}$.

With hindsight, the existence of an operator such as $\mathcal{D}_{p_1 p_2 p_3 p_4}$,
which acts on $\tilde{H}_{2222}$ to give the other correlators, can be understood from the derivative relations between $\bar{D}$ functions.
Note however this way of writing the correlator is not unique,
which is made apparent e.g. by comparing the 4444 cases from \cite{Dolan:2006ec} and \cite{Aprile:2017xsp}.
The fact that the operators $\mathcal{D}_{p_1 p_2 p_3 p_4}$ can be taken
from a unique generating function (\ref{Generating Function 10-4D}) is very surprising and had not been observed before.
It is also completely invisible in the supergravity Feynman rules.

From the perspective of the analytic bootstrap, since the formula (\ref{Generating Function 10-4D}) manifests crossing symmetry,
to rigorously establish it it suffices to show that it correctly predicts the double-discontinuity in one channel, say $v\to 0$.
This will be entirely generated by the $v\to 0$ poles of the seed:
\be
 \dDisc \tilde{H}_{2222} = -\frac{u^3(1+u)}{(1-u)^2v} - \frac{2u^4 \log u}{(1-u)^3 v} + \mbox{pole-free in $v$}.
\ee
Indeed, acting with the differential operators (\ref{D examples}) on this seed we reproduce precisely the double discontinuity
of the finite sum (\ref{Eq:CorrelatordDisc}).  We checked this explicitly for all correlators $p_1,p_2,p_3,p_4$ with $p_i\leq 10$,
making us confident that a group-theoretic identity underwrites the formula (\ref{Generating Function 10-4D}).
It would be interesting to prove it analytically in the general case.

\subsection{Deriving the Mellin-space formula by Rastelli and Zhou} \label{ssec:Rastelli-Zhou}

A frequently useful language for conformal correlators is based on Mellin-space,
see refs.~\cite{Penedones:2010ue, Dolan:2011dv, Gopakumar:2016cpb, Gopakumar:2016wkt} for example.
Correlators of $\mathcal{N}=4$ SYM at strong coupling for arbitrary external scalar operators were conjectured in
ref.~\cite{Rastelli:2016nze,Rastelli:2017udc,Aprile:2018efk} in Mellin-space.  In this section we show that our proposed formula is in precise agreement with that conjecture.

The Mellin representation of the generating functional is an integral over Mandelstam variables such that
\begin{equation}
\tilde{H}_{p_1 p_2 p_3 p_4} = \int \frac{ds dt}{(2\pi i)^2} \ \tilde{M}_{p_1 p_2 p_3 p_4} \label{mellin}
\end{equation}
where the region of integration is detailed in ref.~\cite{Rastelli:2017udc}. We will omit the use of the Mandelstam variable $u$ and express everything in terms of $s,t$ to avoid confusion with the cross-ratio $u$. The reduced correlator for 2222 in position space is given by eq.~\eqref{H2222 Dbar}.
According to the first of eq.~(\ref{D examples}) this function serves as the seed of our generating functional.
We thus start with the 2222 correlator with the normalization of eq.~\eqref{10d four scalars},
written in terms of $\bar{D}_{2,4,2,2}$ in its Mellin representation as written in the Appendix of ref.~\cite{Rastelli:2017udc}:
\begin{equation}
\tilde{M}_{2222} = - u^4 \frac{1}{4} u^{\frac{s}{2}-2} v^{\frac{t}{2}-2} \frac{\Gamma(2-\frac{s}{2})^2 \Gamma(2-\frac{t}{2})^2 \Gamma(\frac{s+t}{2})^2}{\left(1 - \frac{s}{2} \right)\left(1 - \frac{t}{2} \right)\left( \frac{s+t}{2}-1 \right)},
\end{equation}
where $u,v$ are cross-ratio variables. To uplift to a 10-dimensional generating functional, we promote the cross-ratios to be 10-dimensional, normalize according to eq.~\eqref{10d four scalars} and expand in terms of 6-dimensional embedding coordinates similar to eq.~\eqref{Generating Function 10-4D} with the auxiliary contour integral variables $a_i$:
\begin{align}
\tilde{M}_{p_1 p_2 p_3 p_4}&= -2 
\oint \prod_{i=1}^4 \left[\frac{da_i\,a_i^{1-p_i}}{2\pi i} \right] \frac{\Gamma(2-\frac{s}{2})^2 \Gamma(2-\frac{t}{2})^2 \Gamma(\frac{s+t}{2})^2}{\left(-2 + s \right)\left(-2 + t \right)\left( -2+s+t \right)} \frac{(u/\sigma)^{\frac{p_1+p_2}{2}-2}}{(1-a_1 a_3)^4 (1-a_2 a_4)^4} \nonumber \\
& \times \left(u \frac{(1-\frac{\sigma}{u} a_1 a_2)(1- a_3 a_4)}{(1- a_1 a_3)(1-a_2 a_4)} \right)^{\frac{s}{2}-2} \left(v \frac{(1-\frac{\tau}{v} a_2 a_3)(1 - a_1 a_4)}{(1 -a_1 a_3)(1- a_2 a_4)}\right)^{\frac{t}{2}-2}. \label{Mellin Generating Functional}
\end{align}

By construction, this satisfies $\tilde{M}_{2222}=M_{2222}$.
The key feature of this equation is that each of the six $\Gamma(-\alpha)$ functions is naturally paired
with one of the six factors of the form $(1-c  a_i a_j)^{\alpha}$, which originate from the six distances $w_{ij}^2$.
To extract a given spherical harmonics, we need to series-expand in powers of the $a$'s.
Thanks to the pairing, this series can be organized into shifted $\Gamma$-functions
using the identity $-\alpha\Gamma(-\alpha) = \Gamma(-\alpha+1)$:
\begin{equation}
(1-x)^\alpha \Gamma(-\alpha) = \displaystyle \sum_{i=0}^{\infty} \frac{x^i}{i!} \Gamma(i-\alpha).
\label{shift-expansion of gammas}
\end{equation}
Inserting this identity six times into eq.~(\ref{Mellin Generating Functional}), we find a sum of the form 
\be
 \sum_{i,j,k,l,m,n\geq 0} \frac{\Gamma(\ldots)}{i! j! k! l! m! n!}
 \left(\frac{\sigma}{u}\right)^i \left( \frac{\tau}{v}\right)^k
  \left(a_1 a_2\right)^i
  \left(a_3 a_4\right)^j
  \left(a_2 a_3\right)^k
  \left(a_1 a_4\right)^l  \left(a_1 a_3\right)^m  \left(a_2 a_4\right)^n.  \label{six shifts}
\ee
The four contour integrals then select out the terms which have the correct exponent for each $a$.
Consider for example the 3333 case, where each $a$ must appear linearly: three terms contribute,
\begin{align}
\tilde{M}_{3333}&= \frac{-2}{(-2+s)(-2+t)(-2+s+t)} \Big(u^{\frac{s}{2}-2} v^{\frac{t}{2}-2} \Gamma(3-\frac{s}{2})^2 \Gamma(2-\frac{t}{2})^2 \Gamma(\frac{s+t}{2})^2 \nonumber \\
& \quad + u^{\frac{s}{2}-1} v^{\frac{t}{2}-3} \frac{\tau}{\sigma} \Gamma(2-\frac{s}{2})^2 \Gamma(3-\frac{t}{2})^2 \Gamma(\frac{s+t}{2})^2 \nonumber \\
& \quad + u^{\frac{s}{2}-1} v^{\frac{t}{2}-2} \frac{1}{\sigma} \Gamma(2-\frac{s}{2})^2 \Gamma(2-\frac{t}{2})^2 \Gamma(1+\frac{s+t}{2})^2 \Big).
\end{align}
If one shifts $s$ and $t$ for each term above such that there is an overall factor of $u^{\frac{s}{2}-2} v^{\frac{t}{2}-3}$, one obtains
\begin{align}
\tilde{M}_{3333}&= -2u^{\frac{s}{2}-2} v^{\frac{t}{2}-3}  \Gamma(3-\frac{s}{2})^2 \Gamma(3-\frac{t}{2})^2 \Gamma(\frac{s+t-2}{2})^2  \Big( \frac{1}{(-2+s)(-4+t)(-4+s+t)} \nonumber \\
& \quad + \frac{\tau}{\sigma} \frac{1}{(-4+s)(-2+t)(-4+s+t)} + \frac{1}{\sigma} \frac{1}{(-4+s)(-4+t)(-6+s+t)} \Big),
\end{align}
which is precisely the form found in ref.~\cite{Rastelli:2016nze}\footnote{Their convention of $\sigma,\tau$ differs from ours: in their paper, they have
\begin{equation}
\sigma = \frac{y_{13}^2 y_{24}^2}{y_{12}^2 y_{34}^2}, \qquad \tau = \frac{y_{14}^2 y_{23}^2}{y_{12}^2 y_{34}^2}, 
\end{equation}
which should be compared to our definitions given by eq.~\eqref{cross-ratios sigma tau}.
} .
Notice how all the $\Gamma$-functions automatically line up after the shifts.  This can be tracked to the identity
in eq.~(\ref{shift-expansion of gammas}).
In general, one should perform appropriate shifts for each term such that an overall factor of $u^{\frac{s}{2}-2} v^{\frac{t-(p_2+p_3)}{2}}$ can be retrieved.

Let us now derive the general Mellin amplitude from eq.~\eqref{Mellin Generating Functional}.
The Mellin amplitude in \cite{Rastelli:2016nze} was conjectured by solving a set of algebraic crossing-symmetry
constraints.
(These authors assumed, with no loss of generality,
an ordering $p_1 \geq p_2 \geq p_3 \geq p_4$, and ref.~\cite{Aprile:2018efk} also assumed $p_{43}\geq p_{21}\geq 0$, 
but note that here we will make no assumption on the orderings.)
Using eq.~\eqref{six shifts}, the contour integrals in eq.~(\ref{Mellin Generating Functional})
imposes four constraints on the six indices, such that after lining up the powers of $u$ and $v$ we obtain:
\begin{align}
\tilde{M}_{p_1 p_2 p_3 p_4} &= -2 u^{\frac{s}{2}-2} v^{\frac{t-(p_2+p_3)}{2}} \Gamma(\frac{p_1+p_2-s}{2}) \Gamma(\frac{p_3+p_4-s}{2}) \Gamma(\frac{p_2+p_3-t}{2}) \nonumber \\
&\times \Gamma(\frac{p_1+p_4-t}{2}) \Gamma(\frac{s+t+4-p_1-p_3}{2}) \Gamma(\frac{s+t+4-p_2-p_4}{2}) \nonumber \\
&\times \displaystyle \sum_{i,k=0}^{\infty} \frac{1}{i! j! k! l! m! n!} \frac{ \sigma^{i+2-\frac{p_1+p_2}{2}} \tau^k}{(s-2-2s_s)(t-2-t_s)(s+t-2-2(s_s+t_s))}, \label{General Mellin Amplitude}
\end{align}
where the variables above must obey the following constraints:
\begin{align}
j&= \frac{1}{2}(2i -p_1 -p_2 + p_3 + p_4),& l&= \frac{1}{2}(2k+ p_1 -p_2 - p_3 + p_4), \\
m&= \frac{1}{2}(p_1 +p_2 + p_3 -p_4 - 2(2+i+k)), & n&= p_2 - 2 - i -k, \\
s_s&= \frac{p_1 + p_2}{2} - 2 -i ,&t_s&= \frac{p_2+p_3}{2} - 2 - k.
\end{align}
The shifts in Mandelstam variables $s_s,t_s$ were constrained by shifting $s$ and $t$ to obtain the appropriate power of $\frac{p_1+p_2}{2}$ and $\frac{p_2+p_3}{2}$ respectively as explained previously.
Because of the inverse factorials, the summand is only nonvanishing when $i,j,k,l,m,n\geq 0$.
Denoting $i=\frac{p_1+p_2}{2}-2-p$, demanding $m,n\geq 0$ yields for example $k \leq p + \frac{1}{2} \text{min}( p_{21},p_{34})$.
Similarly, one can work out the remaining constraints to find that (this is similar to the summation range for the free correlator
in eq.~\eqref{Gfree Appendix} with $k\mapsto q$):
\begin{align}
p_{\rm min}&= \tfrac12\max(|p_{12}|,|p_{34}|), & p_{\rm max} &=\tfrac12 \min(p_1+p_2,p_3+p_4) - 2,
\\
k_{\rm min}&=\tfrac12\max(0,p_{21}+p_{34}),& k_{\rm max}&=p+\tfrac12\min(p_{21},p_{34}).
\end{align}
This matches precisely the results by ref.~\cite{Rastelli:2016nze}, with some extra factors fixed in ref.~\cite{Aprile:2018efk}.%
\footnote{The precise comparison with ref.~\cite{Aprile:2018efk},
accounting for eq.~(\ref{def Htilde}) and different overall normalizations should be:
\be \tilde{H}^{(1)}_{p_1p_2p_3p_4}=\frac{4}{\sqrt{p_1p_2p_3p_4}}(H_{p_1p_2p_3p_4}^{(1)}-H_{p_1p_2p_3p_4}^{(1){\rm free}})
= \frac{1}{p_1p_2p_3p_4} (N_c^2 H_{p_1p_2p_3p_4})\Big|_{\rm their}.
\ee
Comparing with ref.~\cite{Aprile:2018efk} we thus find precise agreement for the $p_i$ dependence,
up to perhaps an overall constant factor $(4\pi i)^2$ which we didn't track down.}
This completes our derivation of the Mellin amplitude from a more fundamental 10-dimensional symmetry.

\subsection{Ten-dimensional OPE and the leading logarithmic part to all loops}

The SO(10,2) symmetry suggests using ten-dimensional blocks to express the correlators: these should diagonalize the mixing problem.
This will also explain the flat space relation (\ref{Eq:4D-partial-wave}),
which originally motivated our looking for a SO(10,2) symmetry.

We start with the free correlator, eq.~(\ref{Htilde 2222 free}),
and perform its OPE decomposition into 10-dimensional conformal blocks.
As may be seen from the generalized free field formulas from ref.~\cite{Fitzpatrick:2011dm}, an important simplification
occur for fields that saturate the unitarity bound, such as the axi-dilaton:
a \emph{single} double-twist trajectory then contributes,
\be
 \tilde{H}_{2222}^{(0)}(z,\zb) = \sum_{\ell=0,\ {\rm even}}^\infty \tilde{a}^{(0)}_{\ell}\ G^{(10)}_{\ell,8+\ell}(z,\zb)
 \qquad\mbox{with}\qquad a^{(0)}_{(10)}(\ell) = \frac{8\Gamma(4+\ell)^2}{\Gamma(7+2\ell)} (\ell+1)_6.  \label{10D free sum}
\ee
Our blocks are normalized so that $\lim_{z\ll \zb \ll 1}G^{(d)}_{\ell,\Delta}(z,\zb) = z^{\frac{\Delta-\ell}{2}}\zb^{\frac{\Delta+\ell}{2}}$,
and we have verified the sum by series expanding the blocks in $z,\zb$ using ref.~\cite{Hogervorst:2013sma}.
In fact, because we are in even dimensions and on the unitarity bound, the required blocks admit very simple closed-form expressions
as derivatives of hypergeometric functions \cite{Dolan:2011dv}, given explicitly in eq.~(\ref{10D block}).
(Note that since the dimension of the dilaton is protected, the putative 10D CFT cannot be unitary since otherwise this would be the exact
result to all orders in $1/c$.)

To extract ten-dimensional anomalous dimension, we OPE-decompose the $\log u$ terms in
the tree-level correlator $\tilde{H}_{2222}^{(1)}=-u^4 \bar{D}_{2,4,2,2}$:
\ba
 \tilde{H}_{2222}^{(1)}(z,\zb)\Big|_{\log u} &=& -u^4\partial_{u}\partial_v(1+u\partial_u +v\partial_v) \frac{\log(1-z)-\log(1-\zb)}{z-\zb}
\label{H2222 log}
\\
&=& \sum_{\ell=0,\ {\rm even}}^\infty \frac12\tilde{a}^{(0)}_\ell \tilde{\gamma}^{(1)}_\ell\ G^{(10)}_{\ell,8+\ell}(z,\zb)
\quad \mbox{with}\quad \tilde\gamma^{(1)}_\ell \equiv \frac{-1}{(\ell+1)_6}. \label{gamma10 sum}
\ea
This precisely matches the S-matrix phase shift in eq.~(\ref{Eq:10D-partial-wave}).
This agreement has a simple intuitive explanation:
in the bulk point limit $z\to \zb$ (reached after an analytic continuation, see \cite{Heemskerk:2009pn, Maldacena:2015iua} for further detail),
the correlator must reproduce the flat space amplitude.
On the other hand, in this limit, each block in the sum (\ref{gamma10 sum}) reduces to a single Gegenbauer polynomial
matching that in the partial wave expansion (\ref{S matrix partial waves}), weighted by $e^{-i\pi \Delta}$.
Because a single ten-dimensional block appears for each spin, its anomalous dimension
at order $1/c$ must be identical to the flat space phase shift.

The fact that there is a single 10D block for each spin at the lowest order suggests these blocks
``diagonalize'' the double trace mixing problem.
Let us illustrate this with the nontrivial R-symmetry representation $[0,2,0]$ at twist 6, which led to the 2$\times$2 mixing matrix in eq.~(\ref{02 matrix}).
We start by considering the contribution of a single ten-dimensional block to the numerator of the anomalous dimension
matrix in eq.~(\ref{Eq:anomDim}), using the generating function (\ref{Generating Function 10-4D}):
\be
\begin{pmatrix}
\mathcal{D}_{2424} & \mathcal{D}_{2433} \\ \mathcal{D}_{3324}&\mathcal{D}_{3333}
\end{pmatrix}
G^{(10)}_{\ell,8+\ell}(z,\zb), \label{2x2 matrix diff eq}
\ee
where we now project each term onto the $[0,2,0]$ representation using the orthogonality relation (\ref{Eq:Z projection}).
Using the explicit form of the block in eq.~(\ref{10D block}) and expanding at small $z$,
we find that this matrix contains the following three \emph{four-dimensional}
blocks of twist 6 (plus an infinite tower of blocks of twist 8 and higher):\footnote{
There is a slight abuse of notation here as we omit the superscripts on the blocks $G_{\ell,\Delta}^{p_{21},p_{34}}$.
In reality the blocks sit inside the matrix and have a different superscript for each position inside the matrix.}
\begin{align}
&\ell(\ell-1)G_{\ell,\ell+4+6}(z,\zb) \begin{pmatrix}1 & -3/\sqrt{2} \\ -3/\sqrt{2} & 9/2\end{pmatrix}\nonumber
-\frac{\ell(\ell+4)(\ell+7)}{2(\ell+3)}G_{\ell+1,\ell+4+7}(z,\zb)
\begin{pmatrix} 1&0\\0&0\end{pmatrix}\nonumber
\\
&+\frac{(\ell+7)(\ell+8)}{4(2\ell+7)(2\ell+9)} G_{\ell+2,\ell+4+8}(z,\zb) \begin{pmatrix} (\ell+5)^2&3/\sqrt{2}(\ell+4)(\ell+5)\\
3/\sqrt{2}(\ell+4)(\ell+5)&9/2 (\ell+4)^2\end{pmatrix} + O(z^6)
\end{align}
This is not very illuminating yet, but upon shifting the value of $\ell$ in each term so as to extract the coefficient
of $G_{\ell,\ell+10}$ for a given even $\ell$ in the sum of eq.~(\ref{gamma10 sum}), and dividing by the free theory data in the
denominator of eq.~(\ref{Eq:anomDim}), this yields the mixing matrix in the form:
\be
\frac{\gamma^{(1)}_{6,0,2}}{\Delta^{(8)}(4,5+\ell,0,2)}=
 \tilde\gamma^{(1)}_\ell\begin{pmatrix}  \frac{\ell+4}{2\ell+9} & \frac{\sqrt{(\ell+4)(\ell+5)}}{2\ell+9} \\
 \frac{\sqrt{(\ell+4)(\ell+5)}}{2\ell+9} & \frac{\ell+5}{2\ell+9}
 \end{pmatrix}+
  \tilde\gamma^{(1)}_{\ell+2}\begin{pmatrix}  \frac{\ell+5}{2\ell+9} & \frac{-\sqrt{(\ell+4)(\ell+5)}}{2\ell+9} \\
 \frac{-\sqrt{(\ell+4)(\ell+5)}}{2\ell+9} & \frac{\ell+4}{2\ell+9}
 \end{pmatrix}.
\ee
It turns out that each of these matrices is a projector!
The two matrices square to themselves, and are orthogonal to each other.
The right-hand-side precisely reproduces the matrix in eq.~(\ref{02 matrix}), if one inserts
the value of $\tilde\gamma^{(1)}$,  however it is now evident that the eigenvalues are
$\tilde\gamma^{(1)}_\ell$ and $\tilde\gamma^{(1)}_{\ell+2}$ regardless of the functional form of $\tilde\gamma^{(1)}$
used in eq.~(\ref{gamma10 sum}).
That is, the derivative operators in eq.~(\ref{2x2 matrix diff eq}) have simply turned each ten-dimensional block into a projector matrix.

This example shows that the double trace mixing matrices are far from random
but have an entirely group-theoretical origin, which encodes the expansion of SO(10,2) blocks
into those of its SO(4,2)$\times$SO(6) subgroup.  This explains why the eigenvalues are rational numbers,
and also why these numbers are specifically equal to S-matrix phase shifts.

We have verified in many more examples that the ten-dimensional blocks always turn
into projectors onto eigenspaces, including in cases with nontrivial multiplicity.
Presumably, a more thorough study of these matrices would be more easily carried out
at the level of 3-point vertices rather than four-point correlators.

As proposed in ref.~\cite{Aprile:2018efk}, the solution to the double-trace mixing problem at order $1/c$ enables to
compute the leading logarithmic terms at each loop order.
Thanks to the ten-dimensional blocks this can be done for any correlator without actually computing any matrix.
We simply add more powers of $\gamma$ to the ten-dimensional OPE in eq.~(\ref{gamma10 sum}) to find $\tilde{H}_{2222}$,
and then obtain the others by taking derivatives:
\be
 \mathcal{H}^{(k)}_{pqrs}(z,\zb,\a,\ab) \Big|_{\log^{k} u} =
 \left[\Delta^{(8)}\right]^{k-1} \cdot \mathcal{D}_{pqrs} \cdot \mathcal{D}_{(3)}\cdot h^{(k)}(z). \label{conjecture leading log}
\ee
This result is a product of very many differential operators.
The third-order operator $\mathcal{D}_{(3)}$, defined in eq.~(\ref{10D block}), builds
(extremal) ten-dimensional blocks from single-variable functions $h^{(k)}$.
The operator $\mathcal{D}_{pqrs}$, from eq.~(\ref{Generating Function 10-4D}), then extracts various 4D correlators;
for the stress tensor multiplet this operation is trivial, $\mathcal{D}_{2222}=1$.
Finally, the power of $\Delta^{(8)}$, defined in eq.~(\ref{delta8 diff op}),
accounts for the fact that it is only the ratio $\gamma/\Delta^{(8)}$ which is compatible with the ten-dimensional symmetry.
The ordering of operations is important:
in general $\Delta^{(8)}$ destroys the ten-dimensional structure and must act to the left of $\mathcal{D}_{pqrs}$.

Using the explicit form of the block in eq.~(\ref{10D block A}),
together with the coefficients in eqs.~(\ref{10D free sum}) and eqs.~(\ref{gamma10 sum}),
the single-variable seed is given explicitly as:
\be
 h^{(k)}(z)\equiv  \frac{1}{k!}\left(\frac{-1}{2}\right)^k
 \sum_{\ell=0,\ {\rm even}}^\infty \frac{960\Gamma(\ell+1)\Gamma(\ell+4)}{\Gamma(2\ell+7)} \frac{1}{\big[(\ell+1)_6\big]^{k-1}}
z^{\ell+1} {}_2F_1(\ell+1,\ell+4,2\ell+8,z).  \label{llog seed}
\ee
In the first few cases the sum is readily computed,
\ba
  h^{(0)}(z) &=& 2880 \frac{z(2-z)}{1-z}, \\
  h^{(1)}(z) &=& 120\frac{(1-z)^2\log(1-z)}{z^4} + 10\frac{(z-2)(z^2+6z-6)}{z^3} \\
  h^{(2)}(z) &=&  \frac{{\rm Li}_2(z)-(1-z)^5{\rm Li}_2(z/(1-z))}{4z^5} -\frac{(1-z)(2z^2-7z+7)\log(1-z)}{8z^4}
\\ &&  +\frac{(z-2)(1-z)}{z^3}+\frac{235}{576}\frac{z-2}{z},
\ea
with the three-loop case in eq.~(\ref{three loops}).
Plugging $h^{(0)}(z)$ into eq.~(\ref{conjecture leading log}) (and multiplying both sides by $\Delta^{(8)}$) reproduces
eq.~(\ref{Delta8 H0}).  Plugging $h^{(1)}(z)$ we reproduce the logarithmic term given in
eq.~(\ref{H2222 log}) and its extension to all $H_{pqrs}$.
Finally, plugging $h^{(2)}(z)$ we reproduce the one-loop double-discontinuity of $H_{2222}$ computed in refs.~\cite{Aprile:2017bgs,Alday:2017xua,Alday:2017vkk} and now predict the $\log^2u$ terms for all $H_{pqrs}$ correlators.\footnote{
Note that the $\log^2u$ gives the full double-discontinuity at one-loop for $H_{2222}$, but for other correlators
one may also need polar terms. We have not tested whether they are controlled by the same generating function.}
Needless to say, since the one-loop expressions require taking 11 or more derivatives of the compact seed $h^{(2)}(z)$,
these expressions are much lengthier!

\section{Conclusion}

In this paper we revisited the problem of computing correlation functions in holographic conformal field theories,
focusing on the important example of $\mathcal{N}=4$ super Yang-Mills.
We used a conformal field theory approach which constructs OPE data from the singular part of correlators,
the Lorentzian inversion formula \cite{Caron-Huot:2017vep}.
Other approaches to this problem include the direct evaluation of supergravity Witten diagrams
and the solution of crossing symmetry constraints in Mellin-space;
a chief interest of the present method is that the singular part can be computed rather straightforwardly in holographic theories,
and the approach extends seamlessly to one-loop level.

By inserting the identity operator in the cross-channels, we thus obtained the OPE data at order $1/N^0$ in section \ref{Sec:1/c0}; inserting a finite set of protected single-trace operators in section \ref{Sec:Tree-level} then yielded the full $1/N^2$ corrections, corresponding to tree-level supergravity.
In particular, we obtained mixing matrices and anomalous dimensions of double trace operators, see eq.~(\ref{Eq:Anomalous Dimension}),
confirming a recent conjecture \cite{Aprile:2018efk}.

Perhaps our most important result is that an unexpected symmetry controls this mixing matrix.
For one thing, the eigenvalues are rational, as previously observed.
We further observed in section \ref{sec:10D} that they coincide with ten-dimensional flat space partial waves.
We have argued that this agreement can be explained by an accidental SO(10,2) conformal symmetry of the four-particle supergravity amplitude.
This symmetry correctly predicts an infinite number of relation between the correlators of different spherical harmonics, in fact unifying them into a single object. The generating function in eqs.~(\ref{Generating Function 10-4D}) and (\ref{def Htilde})
is the main result of this paper.
It is equivalent to the Mellin-space formula conjectured in \cite{Rastelli:2016nze}, as shown in subsection \ref{ssec:Rastelli-Zhou}.
The symmetry explanation provides further evidence for the validity of this conjecture.
An analytic proof of the symmetry itself, as noted at the end of subsection \ref{ssec:tests},
is now reduced to essentially group-theoretical identities between the singular part of
four and ten-dimensional conformal blocks, which we have verified in many cases.

The SO(10,2) symmetry neatly diagonalizes the mixing between double trace operators
which otherwise carry the same four-dimensional quantum numbers.
Using this, we obtained a compact formula, see eqs.~(\ref{conjecture leading log}) and (\ref{llog seed}), which yields
the leading-logarithmic part of any correlator at each loop order.
(At tree-level, the generating function (\ref{Generating Function 10-4D}) gives more than the leading-log part and
correctly predicts finite parts as well.)  This data
should suffice in the future to determine the one-loop correlator, up to contact ambiguities related to a $R^4$ counter-term,
analogous to \cite{Aprile:2017bgs,Alday:2017vkk}.  An interesting question
is whether the contact ambiguities for different spherical harmonics can be correlated with each other:
this is related to locality in the internal S${}_5$ manifold.

It is very surprising to find that a theory of gravity, which evidently has a dimensionful coupling, effectively behaves like a conformal theory.
Since AdS${}_5\times$S${}_5$ is conformally flat, this relates observables on this geometry to observables in ten-dimensional \emph{flat} space: in our case, anomalous dimensions to $S$-matrix phase shifts.
An analog of this may be the relation between four-dimensional Einstein and conformal gravity, see \cite{Maldacena:2011mk};
conformal-like Ward identities satisfied by tree amplitudes of $n$ gravitons in any dimension were discovered in \cite{Loebbert:2018xce}.

A most pressing question is to determine if such a conformal
symmetry can somehow be formulated for further CFT observables in this or other theories
- higher point correlation functions, higher loops, superstring or M-theory corrections -
or if the symmetry is just an accident of tree-level four-point supergravity amplitudes.
One also wonders if the supergravity Lagrangian could be re-organized to streamline calculations.

\acknowledgments
Work of SCH is supported by the National Science and Engineering Council of Canada
and the Simons Collaboration on the Nonperturbative Bootstrap.
The authors would like to thank Leonardo Rastelli and Xinan Zhou for useful discussions.

\begin{appendix}

\section{Basic inversion integrals} \label{Appendix: Inversion}

Let us record generic results of the inversion integral. It will be convenient to isolate powers of $\frac{\bar{z}}{1-\bar{z}}$ and $\frac{z}{1-z}$ since they integrate nicely. We now perform the $\bar{z}$ integral first as shown in equation 4.7 of \cite{Caron-Huot:2017vep}. First, we compute the double discontinuity using eq.~\eqref{Eq:doubleDisc} to find that
\begin{equation}
\dDisc \left[ \left( \frac{1-\bar{z}}{\bar{z}} \right)^\lambda \bar{z}^{-p_{34}/2} \right] = 2 \sin( \lambda \pi ) \sin( \left( \lambda+ p_{21}+p_{34}\right)\pi ) \left( \frac{1-\bar{z}}{\bar{z}} \right)^\lambda \bar{z}^{-p_{34}/2}.
\end{equation}
To evaluate the $z$ integral, we first rewrite the hypergeometric function in its integral form (with dummy variable $v$) and use the change of variable $t=\frac{\bar{z}(1-v)}{1-\bar{z} v}$. Details of the computation can be found in \cite{Caron-Huot:2017vep}. The $\bar{z}$ integral yields
\begin{align}
\int_0^1 \frac{d\bar{z}}{\bar{z}^2} (1-\bar{z})^{\frac{1}{2}(p_{21}+p_{34})} &\tilde{\kappa} \textstyle \left(\bar{h}/2 \right) \displaystyle k_{\bar{h}/2}^{p_{21},p_{34}}(\bar{z}) \dDisc\left[ \left( \frac{1-\bar{z}}{\bar{z}} \right)^{\lambda} \bar{z}^{-p_{34}/2} \right] \nonumber \\ 
& \qquad = \frac{r_{\bar{h}}^{p_{21},p_{34}}}{\Gamma(-\lambda) \Gamma(-\lambda-\frac{p_{21}+p_{34}}{2})} \frac{\Gamma(\bar{h}-\lambda-\frac{p_{34}}{2} - 1)}{\Gamma(\bar{h}+\lambda+\frac{p_{34}}{2}+1)},
\label{Eq:zbarIntegral}
\end{align}
where $r_{h}^{r,s}$ is defined in eq.~(\ref{rfct}). 

A similar result can be found for the $z$ integral. This integral differs from the previous one by the double discontinuity term which only acted on the $\bar{z}$ sector, and the extra gamma functions from $\tilde{\kappa}$. We therefore have
\begin{align}
\int_0^1& \frac{dz}{z^2} (1-z)^{\frac{1}{2}(p_{21}+p_{34})} k_{1-h}^{p_{21},p_{34}}(z) \left[ \frac{z}{1-z} \right]^\lambda z^{-p_{34}/2} \nonumber \\
=& \frac{\Gamma (2-2 h) \Gamma (1-\lambda ) \Gamma \left(-h+\lambda -\frac{p_{34}}{2}\right) \Gamma \left(\frac{p_{21}}{2}+\frac{p_{34}}{2}-\lambda +1\right)}{\Gamma \left(-h+\frac{p_{21}}{2}+1\right) \Gamma \left(-h-\frac{p_{34}}{2}+1\right) \Gamma \left(-h+\frac{p_{34}}{2}-\lambda +2\right)} \nonumber \\
=& \pi \frac{\sin(\pi (h-\frac{p_{21}}{2})) \sin(\pi(h+\lambda - \frac{p_{34}}{2})) \sin(\pi(h+\frac{p_{34}}{2})) }{\sin(2\pi h) \sin(\pi \lambda) \sin(\pi(h-\lambda+\frac{p_{34}}{2})) \sin(\pi(\frac{p_{21}+p_{34}}{2}-\lambda)) } \nonumber \\
& \times \frac{r_{h}^{p_{21},p_{34}}}{\Gamma(\lambda) \Gamma(\lambda - \frac{p_{21}}{2} - \frac{p_{34}}{2})} \frac{\Gamma(h+\lambda - \frac{p_{34}}{2}-1)}{\Gamma(h-\lambda+\frac{p_{34}}{2}+1)} \nonumber \\
\simeq & \pi \cot(\pi(\lambda-\frac{p_{34}}{2}-h)) \frac{r_{h}^{p_{21},p_{34}}}{\Gamma(\lambda) \Gamma(\lambda - \frac{p_{21}}{2} - \frac{p_{34}}{2})} \frac{\Gamma(h+\lambda - \frac{p_{34}}{2}-1)}{\Gamma(h-\lambda+\frac{p_{34}}{2}+1)}
\label{Eq:zIntegral}
\end{align}
where we have assumed $\lambda>0$ and dropped terms with no poles near positive integer $h$.
Notice the similarity with \eqref{Eq:zbarintegral} setting $\lambda\mapsto\lambda$.

To study higher $1/c$ corrections, one would need to integrate $\left[ \frac{z}{1-z} \right]^\lambda \log(z)$. This will generically yield the following structure:
\begin{align}
\int_0^1& \frac{dz}{z^2} (1-z)^{\frac{1}{2}(p_{21}+p_{34})} k_{1-h}^{p_{21},p_{34}}(z) \left[ \frac{z}{1-z} \right]^\lambda \log(z) z^{-p_{34}/2} \nonumber \\
\simeq & \pi \partial_h \left( \cot(\pi(\lambda-\frac{p_{34}}{2}-h)) \right) \frac{r_{h}^{p_{21},p_{34}}}{\Gamma(\lambda) \Gamma(\lambda - \frac{p_{21}}{2} - \frac{p_{34}}{2})} \frac{\Gamma(h+\lambda - \frac{p_{34}}{2}-1)}{\Gamma(h-\lambda+\frac{p_{34}}{2}+1)} + ... \nonumber \\
\simeq & \frac{1}{\sin^2(\pi h)} \frac{r_{h}^{p_{21},p_{34}}}{\Gamma(\lambda) \Gamma(\lambda - \frac{p_{21}}{2} - \frac{p_{34}}{2})} \frac{\Gamma(h+\lambda - \frac{p_{34}}{2}-1)}{\Gamma(h-\lambda+\frac{p_{34}}{2}+1)} + ...
\label{x log(z) integral}
\end{align}
where the ellipsis correspond to terms with single poles only.
At order $1/c$, the subleading poles would correspond to contributions to OPE coefficients $a^{(1)}$, see \cite{Alday:2017vkk} for example computations.

\section{Superconformal blocks, and weak coupling correlator to order $1/c$} \label{Appendix susy blocks}

Although these were not needed in the main text, for reference we record here expressions
for superconformal blocks, as deduced from the disconnected OPE data of section similarly
to the half-BPS case in eq.~(\ref{B multiplet}).
For the semi-short multiplets $\mathcal{C}_{\ell,m,n}$, equating the Casimirs in eqs.~(\ref{C2 short}) and (\ref{C2 susy})
we find (for generic integer $m,n$) a single matching chiral block:
\be
 f_{\ell+m+2,n-m}(z,\a).
\ee
For the $H$ part, on each of the two families (\ref{protected h}) we find a one-parameter family of solutions, and looking at the
coefficients $\langle a^{(0)} \rangle_{pqqp}$ we find that they all appear with the same coefficient up to sign.
It is thus natural to group them together into a single block:
\be
\mathcal{C}_{\ell,m,n}^{r,s} = \left\{\begin{array}{l}
\displaystyle
k=0, \qquad
f(z,\a) = (-1)^m \ k_{3+\ell+\frac{m+n}{2}}^{r,s}(z)\ k_{-\frac{n-m}{2}}^{-r,-s}(\a),
\\
\displaystyle 
H(z,\zb,\a,\ab) = \sum_{i=1}^{m} (-1)^{i-1}\ G_{\ell+i,6+\ell+m+n-i}^{r,s}(z,\bar{z})\ Z_{m-i,n-i}^{r,s}(\a,\ab) \nonumber \\
 \hspace{25mm}+ \displaystyle \sum_{i=0}^{i_{\rm max}}
 (-1)^{m+i}\ G_{2+\ell+m+i,4+\ell+n-i}^{r,s}(z,\bar{z})\ Z_{i,n-m-i-2}^{r,s}(\a,\ab)
\end{array}\right.
\ee
with $i_{\rm max}=(n-m-2-\max(|r|,|s|))/2$ on the last line.
Again, although deduced from the disconnected correlator, we find that this expression has the correct $z,\zb\to 0$ limit
for all the values of $m,n,\ell,r,s$ that we have verified, and so we identify it to be a general expression for any semi-short block.

We also find that quarter-BPS blocks are simply a special case of the semi-short blocks, where one sets $\ell=-1$:
\begin{equation}
\mathcal{B}_{m,n}^{r,s} = \mathcal{C}_{-1,m-1,n-1}^{r,s}\,. \label{quarter BPS}
\end{equation}
The long block $\mathcal{A}_{\ell,\Delta,m,n}$ only has a $H$ contribution, written already on the last line of (\ref{C2 susy}).

In terms of these superconformal blocks, the OPE decomposition of the disconnected correlator
given in eqs.~(\ref{Eq:a0pqqp})-(\ref{Eq:a0chiral}) turns out to involve exclusively double-traces:
\be\begin{aligned}\label{super OPE disconnected}
\mathcal{G}^{(0)}_{pqqp} &= \delta_{p,q} + \Big(1+\delta_{p,q}\Big)\mathcal{B}_{0,p+q}
\\
& +\sum_{m=0}^{{\rm min}(p,q)-1}\sum_{\ell=-1}^\infty \left[(-1)^m b^{(0)}_{pqqp}(\ell+2+m,p+q-2-2m)\right] \mathcal{C}_{\ell,m,p+q-2-m}
\\
 &+ \sum_{t=0}^\infty \sum_{\ell=0}^\infty\sum_{m,n} a^{(0)}_{pqqp}(\ell,p+q+2t+\ell,m,n)\mathcal{A}_{\ell,p+q+2t+\ell,m,n}\,.
\end{aligned}\ee
The range of $m,n$ is as in eq.~(\ref{small range}) and the coefficients are in eqs.~(\ref{Eq:a0pqqp}) and (\ref{Eq:a0chiral}).
The fact that the OPE data can be reorganized into such a super-OPE with only double trace dimensions
is a non-trivial consistency check.

Let us briefly comment on the so-called multiplet merging phenomenon: there is
an inherent ambiguity in extracting the coefficients of $\mathcal{C}$ blocks,
because a pair of semi-short multiplets can be equivalent to a long one:
\be
\mathcal{C}_{\ell,m,n} + \mathcal{C}_{\ell-1,m+1,n+1} = \mathcal{A}_{\ell,\ell+m+n+2,m,n} \qquad (\ell\geq 0).
\ee
For $\ell=0$, the second term is to be interpreted as a quarter-BPS block according to eq.~(\ref{quarter BPS}).
However, given that all the $\mathcal{C}_{\ell,m,n}$-blocks in eq.~(\ref{super OPE disconnected}) have the maximum value of $m+n$ allowed by R-symmetry considerations, this identity could not be used to rewrite the above without introducing terms with negative coefficients.
We thus conclude that the decomposition of the disconnected correlator into superconformal blocks is unique if we assume unitarity.
More generally, the multiplet merging phenomenon
does not affect the $a$ and $b$ coefficients defined in eqs.~(\ref{Eq:a0pqqp}) and (\ref{Eq:a0chiral}), which are the fundamental objects in this paper.

\begin{figure}[h]
\centering
\def\svgwidth{0.8\linewidth}
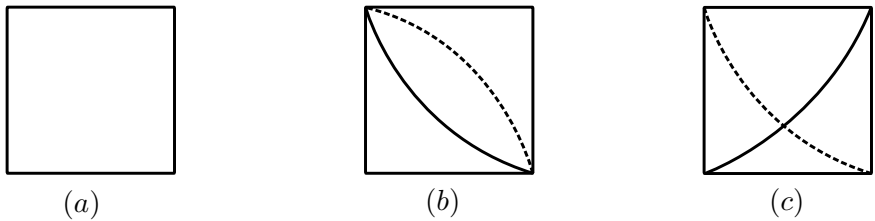
\caption{The three type of planar Wick contractions between four operators at weak coupling;
each line represents a non-empty bundle of propagator.
The respective symmetry factors are 1, $k+1$ and 2, where $k$ is the number of diagonal propagators.
}
\label{Fig:Gfree}
\end{figure}

\subsection{Weak coupling limit}

Finally, let us record another way to look at the protected part of the OPE data.
As discussed in subsection \ref{ssec:free}, due to nonrenormalization theorems and the structure of Wick contractions,
from it one can uniquely determine the \emph{weak} `t Hooft coupling limit.
At order $1/c$, one needs to sum over Wick contractions which can be drawn on a sphere.
There are three cases, as shown in fig.~\ref{Fig:Gfree}; the sum, explicitly, can be written as
\ba
\mathcal{G}^{\rm free}_{\{p_i\}} &=&  \delta_{p_1,p_2}\delta_{p_3,p_4} + \left(\frac{u}{\sigma}\right)^{\frac{p_1+p_2}{2}}
\left[\delta_{p_1,p_3}\delta_{p_2,p_4} + \left(\frac{\tau}{v}\right)^{p_2}\delta_{p_1,p_4}\delta_{p_2,p_3}\right]
\nonumber\\
&& +\frac{\sqrt{p_1p_2p_3p_4}}{4c}
 \sum_{p=p_{\rm min}}^{p_{\rm max}}
\Bigg[
\left(\sum_{q=q_{\rm min}}^{q_{\rm max}} 2\frac{(u/\sigma)^p}{(v/\tau)^q}\right)
+\frac{1+\frac12|p_{23}+p_{14}|}{(v/\tau)^{p+\min(p_{21},p_{34})}}\left(\frac{u}{\sigma}\right)^{\frac12\min(p_1+p_2,p_3+p_4)}
\nonumber\\ && \hspace{35mm}
+\frac{1+\frac12|p_{21}+p_{34}|}{(v/\tau)^{\max(p_{21}+p_{34},0)}}
+\frac{1+\frac12|p_{12}+p_{34}|}{(v/\tau)^{\min(p_{21},p_{34})}}\left(\frac{u \tau}{\sigma v}\right)^p
\Bigg] \nonumber \\ && 
+ O(1/c^2), \label{Gfree Appendix}
\ea
with the summation ranges determined so that all bundles are non-empty,
\begin{align}
p_{\rm min}&=1+\tfrac12\max(|p_{12}|,|p_{34}|),& p_{\rm max}&=\tfrac12\min(p_1+p_2,p_3+p_4)-1, \label{Gfree p constraints}
\\
q_{\rm min}&=1+\tfrac12\max(0,p_{21}+p_{34}),& q_{\rm max}&=p+\tfrac12\min(p_{21},p_{34})-1. \label{Gfree q constraints}
\end{align}
This formula holds when the single-trace operators are defined to be orthogonal to multi-traces,
see discussion below eq.~(\ref{f coeff}).

\section{Ten-dimensional conformal blocks at the unitarity bound}

Some calculations in the text are greatly simplified using the explicit form
of even-dimensional conformal blocks, in particular right at the unitarity bound.
Following arguments in \cite{Dolan:2011dv} we find that these blocks satisfy a simple differential equation:
\be
\left[z\zb \partial_z\partial_{\zb}
+\frac{4}{\zb-z} \left( z^2\partial_z-\zb^2 \partial_\zb\right)\right]  G^{(d=10)}_{\ell,8+\ell}(z,\zb)=0,
\ee
which holds for any $\ell\geq 0$.
By a simple generalization of eq.~(5.10) of \cite{Dolan:2011dv},
a general solution can be written as a differential operator acting on a single-variable function:
\ba
G^{(d=10)}_{\ell,8+\ell}(z,\zb) &=& 
   \Bigg[\left(\frac{z\zb}{\zb-z}\right)^7 f_j(z) +\left(\frac{z\zb}{\zb-z}\right)^6 \frac{z^2}{2}\partial_z f_j(z)
   +\left(\frac{z\zb}{\zb-z}\right)^5 \frac{z^3}{10} \partial_z^2\big(zf_j(z)\big) \nonumber
\\&&\qquad +\left(\frac{z\zb}{\zb-z}\right)^4 \frac{z^4}{120} \partial_z^3\big(z^2f_j(z)\big)\Bigg] +(z\leftrightarrow \zb)
  \equiv \mathcal{D}_{(3)} f_j(z). \label{10D block}
\ea
Comparing with the series expansion in powers of $z,\zb$ (see \cite{Hogervorst:2013sma})
we find that the block indeed takes this form, with the function specifically chosen as
\be
 f_j(z) = \frac{120}{(j+1)(j+2)(j+3)}\ z^{j+1}{}_2F_1(j+1,j+4,2j+8,z). \label{10D block A}
\ee
Finally, for future reference,
let us record the seed controlling the two-loop leading-log term, following eq.~(\ref{llog seed}):
\begin{align}
 h^{(3)} =& \frac{(2-z) (249714 - 249714 z + 217855 z^2)}{12441600 z^3} 
 + \frac{(1 - z)(2903 - 2903 z + 1258 z^2)}{172800 z^4}\log(1-z) 
\nonumber\\ &+\frac{(-2435 + 4870 z - 3670 z^2 + 1235 z^3 + 2 z^4)}{172800 z^4}{\rm Li}_2(z) 
\nonumber\\&  - \frac{(1-z)(227 + 877 z + 177 z^2 - 23 z^3 + 2 z^4)}{345600 z^5}\log^2 (1-z)
\nonumber\\&  + \frac{(1 - z)^3 (1 - 12 z + 21 z^2)}{2880 z^5} \Big(2 g_3(z) + {\rm Li}_3(z)- \log(1-z){\rm Li}_2(z)\Big)
\nonumber\\& +\frac{(1 + 10 z + 10 z^2)}{2880 z^5} \Big(g_3(z)  - {\rm Li}_3(z)\Big),
\label{three loops}
\end{align}
where
$g_3(z) = {\rm Li}_3(1-z)- \log(1-z){\rm Li}_2(1-z) -\frac12\log(z)\log^2(1-z)-\zeta_3 $.

\end{appendix}

\bibliographystyle{JHEP}
\bibliography{references}

\end{document}

%% file: Disconnected-correlators.pdf_tex
\begingroup%
  \makeatletter%
  \providecommand\color[2][]{%
    \errmessage{(Inkscape) Color is used for the text in Inkscape, but the package 'color.sty' is not loaded}%
    \renewcommand\color[2][]{}%
  }%
  \providecommand\transparent[1]{%
    \errmessage{(Inkscape) Transparency is used (non-zero) for the text in Inkscape, but the package 'transparent.sty' is not loaded}%
    \renewcommand\transparent[1]{}%
  }%
  \providecommand\rotatebox[2]{#2}%
  \newcommand*\fsize{\dimexpr\f@size pt\relax}%
  \newcommand*\lineheight[1]{\fontsize{\fsize}{#1\fsize}\selectfont}%
  \ifx\svgwidth\undefined%
    \setlength{\unitlength}{595.27559055bp}%
    \ifx\svgscale\undefined%
      \relax%
    \else%
      \setlength{\unitlength}{\unitlength * \real{\svgscale}}%
    \fi%
  \else%
    \setlength{\unitlength}{\svgwidth}%
  \fi%
  \global\let\svgwidth\undefined%
  \global\let\svgscale\undefined%
  \makeatother%
  \begin{picture}(1,0.26190476)%
    \lineheight{1}%
    \setlength\tabcolsep{0pt}%
    \put(0,0){\includegraphics[width=\unitlength,page=1]{Disconnected-correlators.pdf}}%
  \end{picture}%
\endgroup%

%% file: crossing-symmetry-inversion-formula.pdf_tex
\begingroup%
  \makeatletter%
  \providecommand\color[2][]{%
    \errmessage{(Inkscape) Color is used for the text in Inkscape, but the package 'color.sty' is not loaded}%
    \renewcommand\color[2][]{}%
  }%
  \providecommand\transparent[1]{%
    \errmessage{(Inkscape) Transparency is used (non-zero) for the text in Inkscape, but the package 'transparent.sty' is not loaded}%
    \renewcommand\transparent[1]{}%
  }%
  \providecommand\rotatebox[2]{#2}%
  \newcommand*\fsize{\dimexpr\f@size pt\relax}%
  \newcommand*\lineheight[1]{\fontsize{\fsize}{#1\fsize}\selectfont}%
  \ifx\svgwidth\undefined%
    \setlength{\unitlength}{595.27559055bp}%
    \ifx\svgscale\undefined%
      \relax%
    \else%
      \setlength{\unitlength}{\unitlength * \real{\svgscale}}%
    \fi%
  \else%
    \setlength{\unitlength}{\svgwidth}%
  \fi%
  \global\let\svgwidth\undefined%
  \global\let\svgscale\undefined%
  \makeatother%
  \begin{picture}(1,0.33333333)%
    \lineheight{1}%
    \setlength\tabcolsep{0pt}%
    \put(0,0){\includegraphics[width=\unitlength,page=1]{crossing-symmetry-inversion-formula.pdf}}%
    \put(0.32835423,0.22209664){\color[rgb]{0,0,0}\makebox(0,0)[lt]{\begin{minipage}{0.37960322\unitlength}\raggedright $\displaystyle \int \displaystyle \sum$\end{minipage}}}%
    \put(0.67050787,0.15119033){\color[rgb]{0,0,0}\makebox(0,0)[lt]{\lineheight{1.25}\smash{\begin{tabular}[t]{l}$\displaystyle +$\end{tabular}}}}%
    \put(2.52380952,0.16419198){\color[rgb]{0,0,0}\makebox(0,0)[lt]{\begin{minipage}{0.04761905\unitlength}\raggedright \end{minipage}}}%
    \put(0,0){\includegraphics[width=\unitlength,page=2]{crossing-symmetry-inversion-formula.pdf}}%
    \put(0.01259921,1.10873016){\color[rgb]{0,0,0}\makebox(0,0)[lt]{\begin{minipage}{0.05039682\unitlength}\raggedright \end{minipage}}}%
    \put(-0.00342694,0.20829791){\color[rgb]{0,0,0}\makebox(0,0)[lt]{\begin{minipage}{0.23938492\unitlength}\raggedright $\displaystyle \sum$\end{minipage}}}%
    \put(0.9269416,0.0377976){\color[rgb]{0,0,0}\makebox(0,0)[lt]{\begin{minipage}{0.25198412\unitlength}\raggedright \footnotesize \textbf{Half-BPS}\end{minipage}}}%
    \put(0.64885912,0.06929564){\color[rgb]{0,0,0}\makebox(0,0)[lt]{\begin{minipage}{0.43467261\unitlength}\raggedright \end{minipage}}}%
    \put(0,0){\includegraphics[width=\unitlength,page=3]{crossing-symmetry-inversion-formula.pdf}}%
    \put(0.29878288,0.17063763){\color[rgb]{0,0,0}\makebox(0,0)[lt]{\begin{minipage}{0.22678571\unitlength}\raggedright $\displaystyle =$\end{minipage}}}%
  \end{picture}%
\endgroup%

%% file: ads-flat-space.pdf_tex
\begingroup%
  \makeatletter%
  \providecommand\color[2][]{%
    \errmessage{(Inkscape) Color is used for the text in Inkscape, but the package 'color.sty' is not loaded}%
    \renewcommand\color[2][]{}%
  }%
  \providecommand\transparent[1]{%
    \errmessage{(Inkscape) Transparency is used (non-zero) for the text in Inkscape, but the package 'transparent.sty' is not loaded}%
    \renewcommand\transparent[1]{}%
  }%
  \providecommand\rotatebox[2]{#2}%
  \newcommand*\fsize{\dimexpr\f@size pt\relax}%
  \newcommand*\lineheight[1]{\fontsize{\fsize}{#1\fsize}\selectfont}%
  \ifx\svgwidth\undefined%
    \setlength{\unitlength}{595.27559055bp}%
    \ifx\svgscale\undefined%
      \relax%
    \else%
      \setlength{\unitlength}{\unitlength * \real{\svgscale}}%
    \fi%
  \else%
    \setlength{\unitlength}{\svgwidth}%
  \fi%
  \global\let\svgwidth\undefined%
  \global\let\svgscale\undefined%
  \makeatother%
  \begin{picture}(1,0.47619048)%
    \lineheight{1}%
    \setlength\tabcolsep{0pt}%
    \put(0,0){\includegraphics[width=\unitlength,page=1]{ads-flat-space.pdf}}%
    \put(0.18863215,0.03702532){\color[rgb]{0,0,0}\makebox(0,0)[lt]{\begin{minipage}{0.31454653\unitlength}\raggedright Ads$_5$\end{minipage}}}%
    \put(0.38426368,0.35439152){\color[rgb]{0,0,0}\makebox(0,0)[lt]{\begin{minipage}{0.20826905\unitlength}\raggedright S$_5$\end{minipage}}}%
    \put(0.80645575,0.31852978){\color[rgb]{0,0,0}\makebox(0,0)[lt]{\begin{minipage}{0.20952899\unitlength}\raggedright $\mathbb{R}^{10}$\end{minipage}}}%
    \put(0,0){\includegraphics[width=\unitlength,page=2]{ads-flat-space.pdf}}%
    \put(0.71500496,0.20347718){\color[rgb]{0,0,0}\makebox(0,0)[lt]{\begin{minipage}{0.3143502\unitlength}\raggedright $G_{10}$\end{minipage}}}%
    \put(0.1556002,0.20977678){\color[rgb]{0,0,0}\makebox(0,0)[lt]{\begin{minipage}{0.4731002\unitlength}\raggedright $\mathcal{D}_{p_1 p_2 p_3 p_4} G_{10}$\end{minipage}}}%
    \put(0.64381945,0.30553075){\color[rgb]{0,0,0}\makebox(0,0)[lt]{\begin{minipage}{0.57767361\unitlength}\raggedright $\mathcal{O}^4$\end{minipage}}}%
    \put(0.91222683,0.29591577){\color[rgb]{0,0,0}\makebox(0,0)[lt]{\begin{minipage}{0.57767361\unitlength}\raggedright $\mathcal{O}^4$\end{minipage}}}%
    \put(0.8252923,0.17181359){\color[rgb]{0,0,0}\makebox(0,0)[lt]{\begin{minipage}{0.57767361\unitlength}\raggedright $\mathcal{O}^4$\end{minipage}}}%
    \put(0.56007901,0.16299415){\color[rgb]{0,0,0}\makebox(0,0)[lt]{\begin{minipage}{0.57767361\unitlength}\raggedright $\mathcal{O}^4$\end{minipage}}}%
    \put(0.06430024,0.29780566){\color[rgb]{0,0,0}\makebox(0,0)[lt]{\begin{minipage}{0.25009425\unitlength}\raggedright $\mathcal{O}^{p_2}$\end{minipage}}}%
    \put(0.06274504,0.17884158){\color[rgb]{0,0,0}\makebox(0,0)[lt]{\begin{minipage}{0.25009425\unitlength}\raggedright $\mathcal{O}^{p_1}$\end{minipage}}}%
    \put(0.34622717,0.30042392){\color[rgb]{0,0,0}\makebox(0,0)[lt]{\begin{minipage}{0.25009425\unitlength}\raggedright $\mathcal{O}^{p_3}$\end{minipage}}}%
    \put(0.34559721,0.1738019){\color[rgb]{0,0,0}\makebox(0,0)[lt]{\begin{minipage}{0.25009425\unitlength}\raggedright $\mathcal{O}^{p_4}$\end{minipage}}}%
  \end{picture}%
\endgroup%

%% file: Gfree.pdf_tex
\begingroup%
  \makeatletter%
  \providecommand\color[2][]{%
    \errmessage{(Inkscape) Color is used for the text in Inkscape, but the package 'color.sty' is not loaded}%
    \renewcommand\color[2][]{}%
  }%
  \providecommand\transparent[1]{%
    \errmessage{(Inkscape) Transparency is used (non-zero) for the text in Inkscape, but the package 'transparent.sty' is not loaded}%
    \renewcommand\transparent[1]{}%
  }%
  \providecommand\rotatebox[2]{#2}%
  \newcommand*\fsize{\dimexpr\f@size pt\relax}%
  \newcommand*\lineheight[1]{\fontsize{\fsize}{#1\fsize}\selectfont}%
  \ifx\svgwidth\undefined%
    \setlength{\unitlength}{439.37007874bp}%
    \ifx\svgscale\undefined%
      \relax%
    \else%
      \setlength{\unitlength}{\unitlength * \real{\svgscale}}%
    \fi%
  \else%
    \setlength{\unitlength}{\svgwidth}%
  \fi%
  \global\let\svgwidth\undefined%
  \global\let\svgscale\undefined%
  \makeatother%
  \begin{picture}(1,0.32258065)%
    \lineheight{1}%
    \setlength\tabcolsep{0pt}%
    \put(0.40967742,0.53713259){\color[rgb]{0,0,0}\makebox(0,0)[lt]{\begin{minipage}{0.23670251\unitlength}\raggedright  \end{minipage}}}%
    \put(0.09536648,0.06728323){\color[rgb]{0,0,0}\makebox(0,0)[lt]{\begin{minipage}{0.30043011\unitlength}\raggedright $(a)$\end{minipage}}}%
    \put(0.48978022,0.06903745){\color[rgb]{0,0,0}\makebox(0,0)[lt]{\begin{minipage}{0.22307671\unitlength}\raggedright $(b)$\end{minipage}}}%
    \put(0.86407308,0.0688406){\color[rgb]{0,0,0}\makebox(0,0)[lt]{\begin{minipage}{0.26401434\unitlength}\raggedright $(c)$\end{minipage}}}%
    \put(0,0){\includegraphics[width=\unitlength,page=1]{Gfree.pdf}}%
    \put(0.35505376,1.79053762){\color[rgb]{0,0,0}\makebox(0,0)[lt]{\begin{minipage}{0.22759857\unitlength}\raggedright \end{minipage}}}%
  \end{picture}%
\endgroup%